\documentclass[12pt,preprint]{aastex}

\slugcomment{Submitted to the Astron. J.}

\shorttitle{Chaos and Effects of Planetary Migration on Kiviuq}
\shortauthors{Carruba et al.}

\begin{document}

\title{Chaos and the Effects of Planetary Migration 
on the Orbit of S/2000 S5 Kiviuq}

\author{V. Carruba\altaffilmark{1}}
\affil{Astronomy Department, Cornell University,
    Ithaca, NY 14853, U.S.A.}
\email{valerio@astro.cornell.edu}

\author{D. Nesvorn\'{y}}
\affil{Southwest Research Institute, Boulder, C0 80302, U.S.A.}

\author{J. A. Burns and M. \'{C}uk}
\affil{Astronomy Department, Cornell University,
    Ithaca, NY 14853, U.S.A.}

\author{K. Tsiganis}
\affil{Observatoire de la C\^{o}te D' Azur, B. P. 4229, Nice Cedex 4, France}

\altaffiltext{1}{Present address: Instituto de Astronomia, Geof\'{i}sica e 
Ci\^{e}ncias Atmosf\'{e}ricas, Universidade de S\~{a}o Paulo, CEP
05508-900, S\~{a}o Paulo, SP, Brasil.}
\begin{abstract}

Among the many new irregular satellites that have been discovered 
in the last five years, at least six are in the so-called 
Kozai resonance.  Due to solar perturbations, 
the argument of pericenter $\omega$ of a satellite usually precesses 
from 0 to $360^{\circ}$.  However, at inclinations higher than 
$\simeq 39.3^{\circ}$ and lower than $\simeq 140.7^{\circ}$ a new 
kind of behavior occurs for which 
the argument of pericenter oscillates around $\pm 90^{\circ}$.  In 
this work we will concentrate on the orbital history of 
the saturnian satellite S/2000 S5 Kiviuq, one of the satellites currently 
known to be in such resonance

Kiviuq's orbit is very close to the separatrix of the Kozai resonance.
Due to perturbations from the other jovian planets, it is expected that orbits
near the Kozai separatrix may show significant chaotic behavior.  This 
is important because chaotic diffusion may transfer orbits from libration 
to circulation, and vice versa.  To identify 
chaotic orbits we used two well-known 
methods: the Frequency Analysis Method 
(Laskar 1990) and Maximum Lyapunov Exponents (Benettin {\em et al.} 1980).
Our results show that the Kozai resonance is crossed by a web of secondary 
resonances, whose arguments involve combinations of 
the argument of pericenter, the argument of the Great Inequality 
($2{\lambda}_{Jupiter}-5{\lambda}_{Saturn}$), longitude of the 
node $\Omega$, and other 
terms related to the secular frequencies $g_5$, $g_6,$ and $s_6$.  
Many test orbits whose precession period is close 
to the period of the Great Inequality (883 yrs), or some of its harmonics, 
are trapped by these secondary resonances, and show 
significant chaotic behavior.

Because the Great-Inequality's period is connected to the semimajor axes 
of Jupiter and Saturn, and because the positions of the jovian planets have   
likely changed since their formation (Malhotra 1995), the phase-space 
location of these secondary resonances should have been different in 
the past.  By simulating the effect of planetary migration, we show that 
a mechanism of sweeping secondary resonances, similar to the one 
studied by Ferraz-Mello {\em et al.} (1998) for the 
asteroids in the 2:1 mean motion 
resonance with Jupiter, could significantly deplete 
a primordial population 
of Kozai resonators and push several circulators near the Kozai separatrix.
This mechanism is not limited 
to Kiviuq's region, and could have worked to destabilize any  
initial population of satellites in the Kozai resonance 
around Saturn and Jupiter.

\end{abstract}

\keywords{celestial mechanics, satellites, resonances}

\newpage
\section{Introduction}
 
Carruba {\em et al.} (2002) studied an analytical model of the 
Kozai resonance (Kozai 1962) 
based on a secular development of the disturbing function, 
expanded in series of $a/a'$ (where the $'$ stands hereafter for quantities 
related to the perturber), truncated at second order in $a/a'$, and averaged 
over the mean anomalies of both perturber and perturbee (see also 
Innanen {\em et al.} 1997).
Two kinds of behavior are possible in this simplified model: for  
inclinations of less than 39.23$^{\circ}$ (or higher than 140.77$^{\circ}$ 
for retrograde satellites), the argument of pericenter freely circulates from 
0$^{\circ}$ to 360$^{\circ}$, while at intermediate inclinations a new class 
of solutions, in which the argument of pericenter librates around 
$\pm$90$^{\circ}$, are possible.  Such behavior is called the Kozai 
resonance.
The value of the critical inclination of 39.23$^{\circ}$ is, in fact, 
an artifact of the averaging; the real boundary between the 
region in which libration is possible or not is actually 
a more complicated 
function of the satellite's semimajor axis and eccentricity 
(\'{C}uk {\em et al.} 2003, \'{C}uk and Burns 2004). 
Currently, six irregular satellites of jovian planets are known 
numerically to be in the 
Kozai secular resonance: two around Saturn (S/2000 S5 Kiviuq, and 
S/2000 S6 Ijiraq), two around Jupiter 
(S/2001 J10 Euporie, 
and S/2003 J20), and 
two around Neptune 
(S/2002 N2 and S/2002 N4, Holman {\em et al.} 2004).  A third neptunian 
satellite, S/2003 N1 is on a chaotic orbit that may also be in the 
Kozai resonance (Holman {\em et al.} 2004).  

A question that Carruba {\em et al.} left unanswered regarded the possible 
presence of chaos and chaotic diffusion 
at the separatrix between the regions of circulation and libration.  
Since in that work the authors considered a three-body 
problem, but averaged over the mean anomalies of both 
perturber and perturbee, our analytical model was an 
integrable one-degree of freedom model.
In the real system, however, perturbations from
other jovian planets and short-period terms may produce chaos
near the Kozai separatrix.   
Chaotic evolution might lead to the escape
of objects whose orbits were originally inside the Kozai resonance.
Here we test the hypothesis that the chaotic layer is due to 
the fact that the Kozai resonance overlaps with secondary resonances 
near the separatrix.  It is well known that resonance overlap is 
one of the major sources of chaos in dynamical systems (Chirikov 1979).
Moreover, the positions of the 
planets may have changed after their formation, due to gravitational  
scattering of planetesimals (Malhotra 1995). 
As a consequence of the different planetary positions, the shape of 
the chaotic layer and the locations of the 
secondary resonances inside and outside the Kozai resonance should have 
been different
in the past.  This might have had consequences on the stability
of any primordial population of Kozai resonators.

This paper will concentrate on identifying the chaotic layer
at the transition between circulation and 
libration for Kozai resonators, in particular for the 
saturnian satellite Kiviuq, whose orbit is very 
close to the separatrix of the Kozai resonance.  Our goal is 
to understand the origins of chaos and the effects of 
planetary migration.

Our investigation is divided in the following way:  section 2 describes the major features of our simplified model of the 
Kozai resonance and identifies the transition region between 
circulation and libration.  Section 3 recalls
two methods to identify chaotic 
behavior in dynamical systems, the Frequency Analysis Method 
and the maximum Lyapunov exponents.  Section 4 investigates possible causes 
of the chaotic layer.  Section 5 shows how the Kozai resonance is
crossed by a series of secondary resonances of different strengths.  In the final section we demonstrate how the effect of planetary migration,
combined with the effect of the secondary resonances, 
might have operated to depopulate a possible 
primordial satellite group of Kozai resonators. 

\section{Identifying the Transition between 
Circulation and Libration: Analytical and Numerical Tools}

Carruba {\em et al.} (2002) described a simple 
analytical model of the Kozai resonance.
This model is based on a development of the disturbing function in 
a series of $r/r'$, where $r$ and $r'$ are the 
radial planetocentric distances of the satellite and 
the Sun, truncated at the second order.  The perturbing 
function $R$ is:

\begin{equation}
R=\frac{G_0 m'}{r'} P_{2}(s)\left(\frac{r}{r'}\right)^2,
\label{eq: 4_1}
\end{equation}

\noindent where $G_0$ is the gravitational constant, $m'$ is 
the solar mass, $s$ is 
the angle between directions to the satellite and the Sun as seen from 
the planet, and 
$P_{2}$ is the second-order Legendre polynomial. 
The perturbing function, when averaged over the mean anomalies
of both Sun and satellite, becomes:

\begin{equation}
R=\frac{G_0 m'a^2}{8b'^3}\left[2+3e^2-\left(3+12e^2-
15e^2{\cos^2{\omega}}\right){\sin^2{I}}\right],
\label{eq: 4_2}
\end{equation}

\noindent where $a, e,$ and $I$ are the semimajor axis, eccentricity, and 
inclination of the satellite, respectively.  $b'= a'\sqrt{1-e'^2}$ is 
the Sun's semiminor axis.  
The perturbing function, when expressed in terms of Delaunay 
elements ($l,g,h,L,G,H$; see Murray and Dermott 1999, Danby 1988), is:

\begin{equation}
R=-\frac{m'L^4}{8b'^3 G_0 m'}\left[-10+\frac{3}{L^2} (3G^2 - 4H^2) 
+\frac{15H^2}{G^2}+15{\cos^2{g}}\left(1-\frac{G^2}{L^2}-\frac{H^2}{G^2}
+\frac{H^2}{L^2}\right) \right].
\label{eq: 4_3}
\end{equation}

\noindent Since both $l$ and $h$ are cyclic, the 
conjugate quantities $L$ and $H$ 
are constants of the motion, as is the 
quantity:

\begin{equation}
\Theta=\left(1-e^2\right){\cos^2{I}}=H^2/L^2.  
\label{eq: Theta}
\end{equation}

\begin{figure}
\epsscale{1.10}
\plottwo{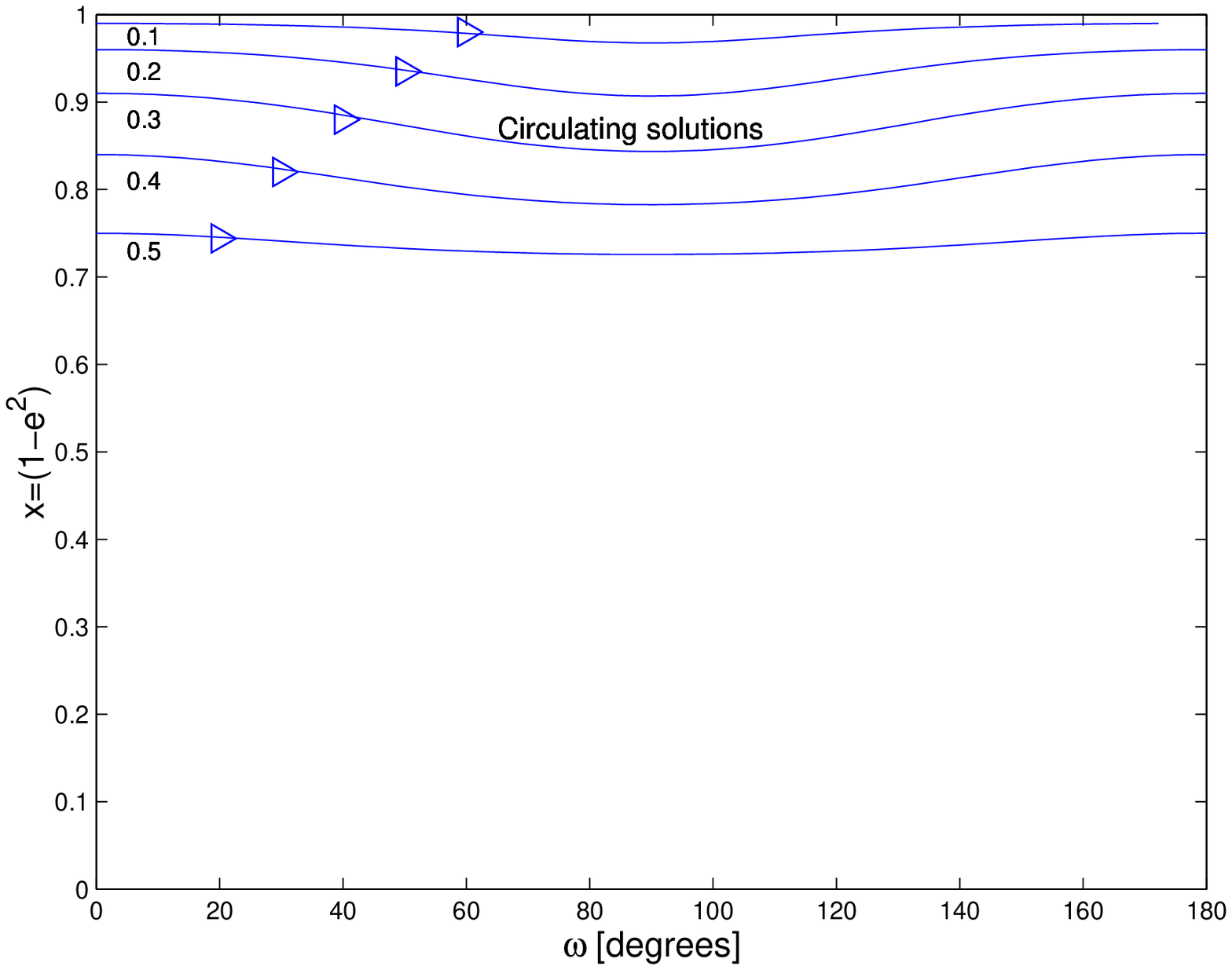}{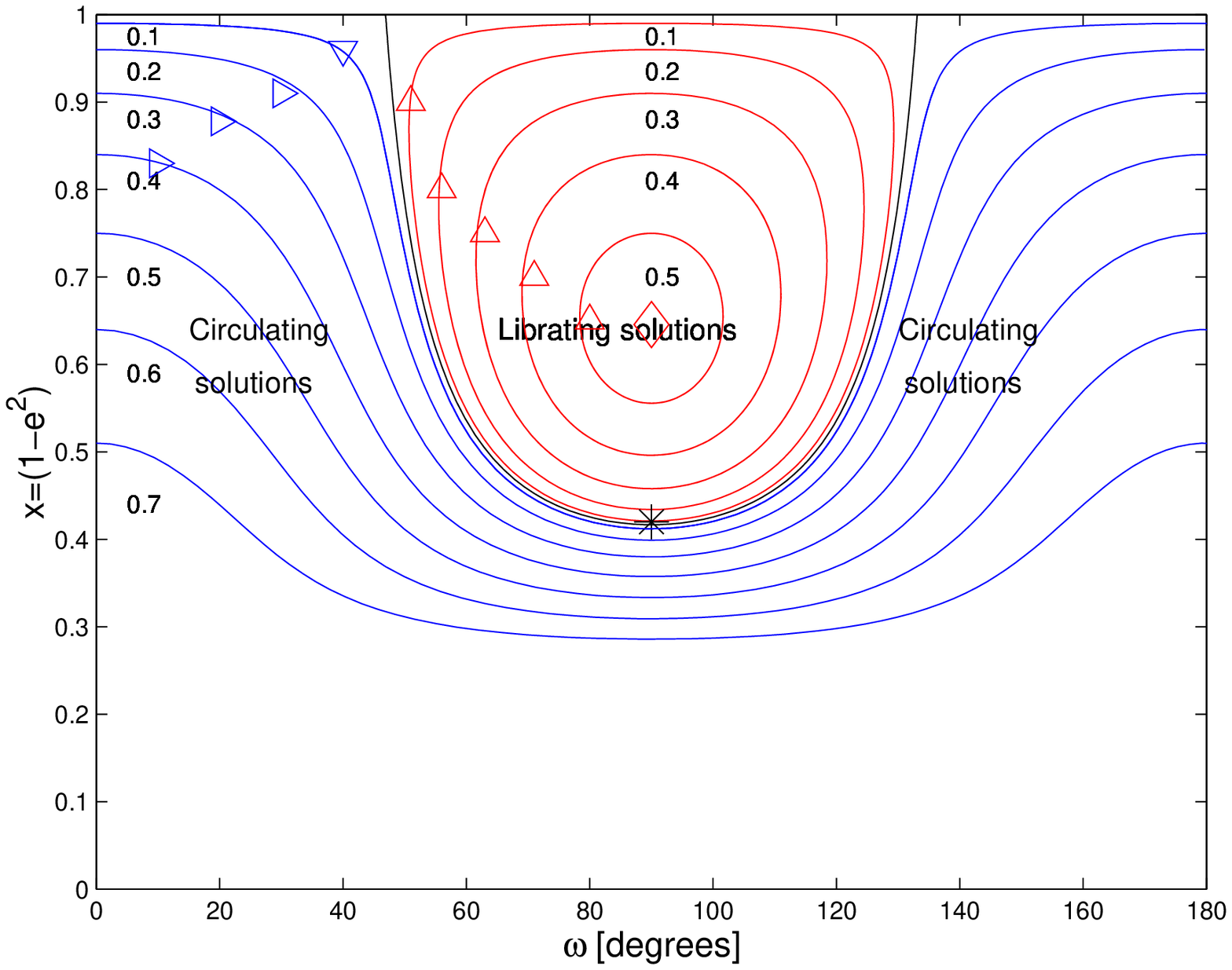}
\caption{Energy levels of the Kozai Hamiltonian for 
$\Theta$ = 0.70 and = 0.25, in the $\omega$ vs. $1-e^2$ plane.  Since the 
levels are symmetric about $\omega$ = $180 ^{\circ}$, 
we only show the interval $0-180^{\circ}$.  For 
$\Theta < 0.6$ librating solutions are possible 
(see the right panel).} 
\label{fig: c_fig1}
\end{figure}

\noindent With $L$ and $H$ fixed, the perturbing function (\ref{eq: 4_3}) 
represents a one-degree of freedom model for the Kozai resonance
(Innanen {\em et al.} 1997).
The following equations for the orbital elements hold:

\begin{eqnarray}
\frac{d I}{d \tau} & = & -\frac{15}{16\sqrt{1-e^2}} 
\left(e^2\sin{2\omega}\sin{2I}\right), \\
\frac{d e}{d \tau} & = & \frac{15}{8}e\sqrt{1-e^2}\sin{2\omega}~{\sin^2{I}}, \\
\frac{d \omega}{d \tau} & = & \frac{3}{4\sqrt{1-e^2}}
\left[2(1-e^2)+5{\sin^2{\omega}}(e^2-{\sin^2{I}})\right],
\end{eqnarray}

\noindent where $\tau={G_0 m't}/{nb'^3}$ and t is time.  
The time has been rescaled in this way to eliminate
the equations' dependency on the satellite's semimajor axis.  
Kinoshita and Nakai (1999) found the solution of (5-7) in terms of elliptic 
functions.  Fig.~\ref{fig: c_fig1} shows the results of 
numerical integrations of equations (5-7) for 
$\Theta=0.70$ and $\Theta=0.25$.  For $\Theta$ less than 0.6, librating 
solutions are possible.

This model has several limitations.  First, 
short-period terms have been eliminated by averaging.  One 
consequence of this averaging is that the limit between the zones in 
which libration is possible
or not is $I = 39.23^{\circ}$ (or $140.77^{\circ}$ for 
retrograde orbits), 
regardless of the satellite's semimajor
axis (both inclinations correspond to $\Theta$ = 0.6 for $e = 0$).  When short-period terms are included, the boundary 
shifts slightly (\'{C}uk and Burns 2004).  A 
case where this is observed is the orbit of the jovian satellite 
Euporie that is a Kozai resonator but has $\Theta > 0.6$.

A second limitation of the averaged model is that, since 
it is an integrable, 
one-degree-of-freedom model, no chaos can appear at the separatrix between 
circulation and libration.  Finally, since 
mutual perturbations among jovian planets are ignored,
the orbital elements of the planet about which the 
satellites circle are constant.  In the real solar system, the 
eccentricities of Jupiter and Saturn vary with several prominent periods 
associated with the secular frequencies $g_5$ and $g_6$ (timescales 
of up to $1/g_5 \simeq300,000$ yr), Great-Inequality terms $2\lambda_{J}-
5\lambda_{S}$, where the suffices J and S hereafter stand 
for Jupiter and Saturn, (with a timescale of 883 yr), and 
short-period frequencies, 
connected with the orbital periods of the planets (12.5 and 29.7 yrs 
for Jupiter and Saturn, respectively).  The shape of the separatrix 
may change in time according to the variations of the planet's  
eccentricity and semimajor 
axis (remember the presence of the factor ${b'}^3$ in $\tau$) 
which might introduce chaos along the real system's separatrix.

To understand the behavior of the real system, we 
concentrate on the case of Kiviuq, a saturnian satellite in Kozai 
resonance whose orbit is very close to the separatrix 
between libration and circulation in $\omega$.  We start 
by numerically searching for the transition 
between circulation and libration.  To do so, we used the following 
procedure: we numerically integrated Kiviuq's orbit over 100,000 yrs and found 
the maximum eccentricity, because then the 
satellite is closest to the separatrix.
We computed $\Theta$ according to the inclination of Kiviuq 
at that time, and generated a grid of initial conditions for test 
particles; this contains 19 values in $\omega$, from $40^{\circ}$ to 
$140^{\circ}$, 
separated by $5^{\circ}$.  All particles were given the average 
value of Kiviuq's semimajor axis during the integration ($a=0.0755$ AU), and 
a grid in inclination with 21 values, starting at $38.6^{\circ}$ with a step 
of $0.6^{\circ}$.  The values of the eccentricities were computed from 
$\Theta = 0.4464$.  Initial $\Omega$ and $M$ (mean anomaly)
are equal to those of Kiviuq at its maximum eccentricity.  
This grid forms our ``low-resolution survey''. 

These initial orbits were integrated for 2 Myr under the influence of the four 
jovian planets (Outer Solar System, OSS hereafter) with a modified 
version of SWIFT-WHM, the 
integrator that uses the Wisdom and Holman (1991) mapping in the SWIFT 
package (Levison and Duncan 2000).  We modified the integrator so 
that the planets' orbits are integrated in the heliocentric system and 
the particles' orbits are integrated in the planetocentric reference frame.
This setup minimizes the integration error (Nesvorn\'{y} {\em et al.} 2003).

\begin{figure}
\epsscale{1.01}
\plotone{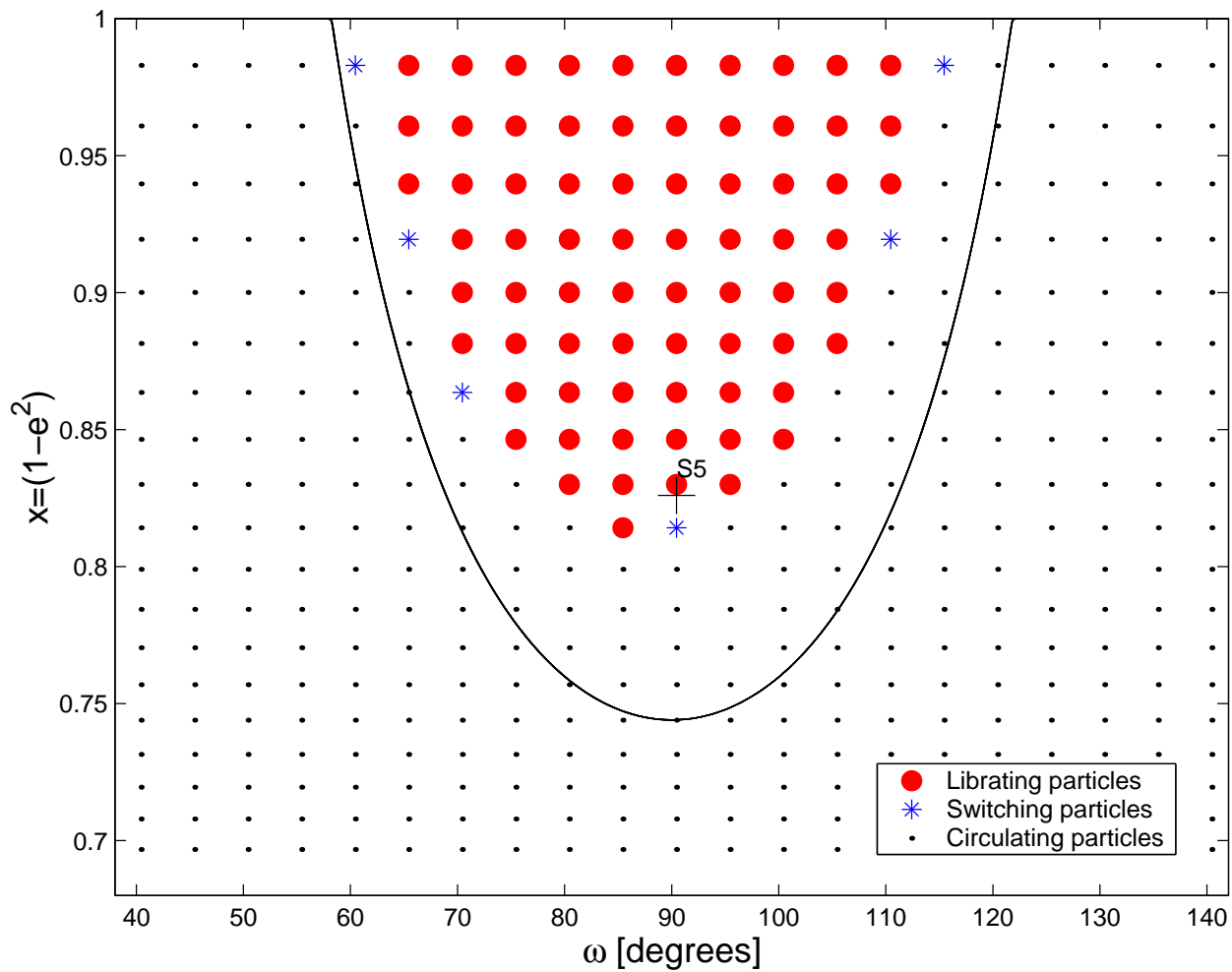}
\caption{The fate of orbits located near the separatrix 
between circulation and libration for our low-resolution survey.  
Librating particles are shown by full dots, 
circulating orbits are shown by black points; and asterisks 
denote orbits  
that switched from libration to circulation.
The black line identifies the separatrix according to the secular model for 
S/2000 S5 Kiviuq's $\Theta$ at the maximum eccentricity.
Kiviuq's position is indicated by a cross (S5 in the figure).}
\label{fig: particles_fate}
\end{figure}

To more easily identify secular oscillations in Saturn's inclination, 
we refer all the positions to Saturn's initial orbital plane.
Fig.~\ref{fig: particles_fate} shows the results of our 
simulations for our low-resolution survey.  The actual separation
between circulating and librating behavior does not follow the solution
of the secular model when the full perturbations from jovian planets are 
considered.  Moreover, some orbits alternate between circulation and 
libration (and, in a few cases, spend some time in the other libration 
island around $\omega = 270^{\circ}$).

The question that remains to be answered is whether there is a chaotic 
layer near the separatrix, and if chaotic diffusion might be effective in 
depleting a primordial population of Kozai resonators.  
To address this question, and identify chaos, 
we will use the Frequency 
Analysis Method (FAM, hereafter; Laskar, 1990), and will compute the Maximum 
Lyapunov Exponents (MLE hereafter; Benettin {\em et al.}, 1980) which
we now describe.

\section{Numerical Tools to Study Chaotic Orbits}

\begin{figure}
\plotone{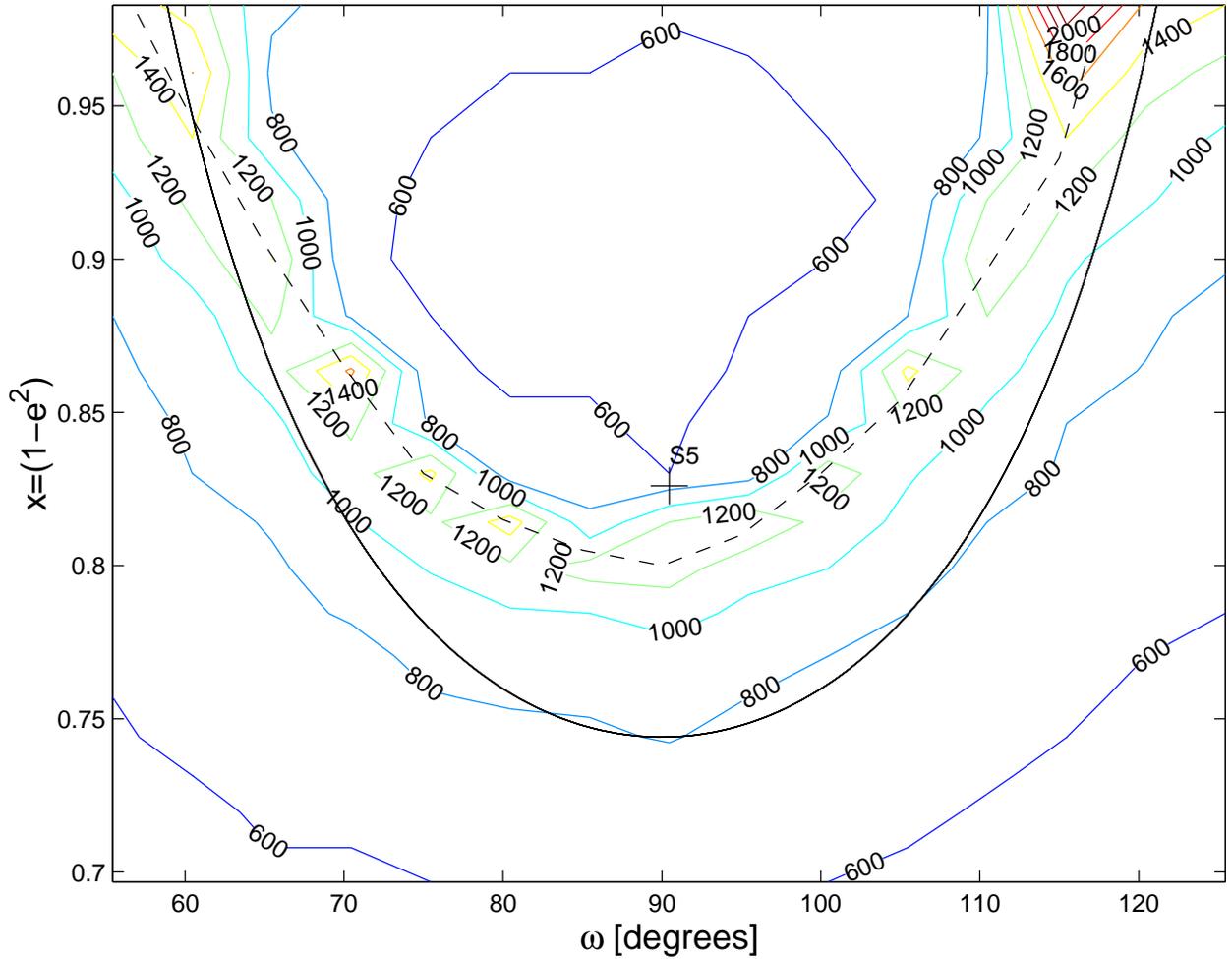}
\caption{Contour plots that show the periods of libration or 
circulation in $\omega$ in years.  The 
argument of pericenter $\omega$) lies on the horizontal axis, and 
on the vertical axis $x=1-e^2$.  The solid parabolic line 
identifies the separatrix according to the secular model for 
S/2000 S5 Kiviuq's $\Theta$ at the maximum eccentricity.  The dashed line
shows the approximate location of the actual separatrix 
(see Fig.~\ref{fig: particles_fate}).
\label{fig: Periods}}
\end{figure}

The Frequency Analysis Method (FAM) is essentially an analysis of the frequency power spectrum of appropriate
combinations of numerically determined orbital elements.  In our case, 
for example, we are interested in the frequency related to the 
precession of the argument of the pericenter $\omega$.  
Fig.~\ref{fig: Periods} shows the periods 
in $\omega$ for the particles in 
our low-resolution survey.  The periods peak 
sharply in the transition region from circulation to libration in 
$\omega$, reaching a maximum of 1800 yrs. The minimum libration period is 
480 yrs.  The 
average period of libration in the Kozai resonance in our survey is 640 yrs.
This range in periods corresponds
to a frequency range of $\simeq$ 700-3200~``/yr.

We took the Fourier 
transform of $e \cdot \exp{\iota \omega}=  e \cos{\omega}+
\iota e \sin{\omega}$ (where $\iota=\sqrt{-1}$) over two time 
intervals (0-1 Myr and 1-2 Myr) and computed the quantity:

\begin{equation}
\sigma=\log{\left| 1-\frac{f^{2}}{f^{1}}\right|},
\end{equation}

\noindent where $f^{1}$ is the frequency over the first interval and 
$f^{2}$ the corresponding frequency on the second interval.  If 
the motion is regular, $f(t)$ will be nearly constant, with small 
variations produced by errors 
in the determination of the frequencies 
(to be discussed later in this section).  If the motion is chaotic, 
variations in $\sigma$ will be related to changes in 
$\omega$'s precession period.

While this is the main idea behind the FAM, a few complications occur.  
First, the precision of a Fast Fourier Transform (FFT hereafter)
is limited by its coarse resolution [$\Delta f=\frac{1}{N \Delta \tau}$, where
$\Delta \tau$ is the (constant) time separation of $\omega$-sampling 
and N is the number of data points used, which must be a power of 
2 (Press {\em et al.} 1996)] 
in recovering the frequencies having the largest amplitude.
To overcome this problem, we used the Frequency Modified Fourier 
Transform method (FMFT), with quadratic corrections (\v{S}idlichovsky 
and Nesvorn\'{y} 1997), which allows us to retrieve the frequencies 
with the largest amplitude to a better precision.  To estimate the 
error associated with this method, we computed the frequencies 
with and without quadratic corrections, and computed the average 
value of the differences in $\sigma$.
This procedure gives an upper limit on the error.
In our case, the average error of FMFT was $\sigma=-4.5$, while 
a simple FFT had a resolution of $\sigma=-3$ (we 
used a time step $\Delta \tau$=30 yr, and 
32,768 data points, so that each interval for the determination of 
the frequencies was $\simeq$ 1~Myr).

Another problem to take into account is {\it aliasing}.  The FFT can 
recover frequencies up to the Nyquist frequency 
($f_{N}=\frac{1}{2\Delta \tau}$, 
that corresponds to two samplings per period).  Frequencies higher than the
Nyquist frequency are not lost, but are mapped into a 
smaller apparent value of $f$, given by $f_{app}=f_{N}-(f_{real}-f_{N})$.
This phenomenon is called aliasing and 
can be easily understood by considering a sinusoidal wave, sampled 
at isolated points: at least two points per period are needed 
to obtain a good estimate of the period .   If the sampling is longer than 
one period, than the retrieved period is larger, and its ``ghost'' signature 
may appear in the frequency range of interest.

To avoid this problem, we used an ``on-line'' low-pass filter, following 
the procedure of Quinn {\em et al.} (1991) (based on the work 
of Carpino {\em et al.} 1987).  Our filter suppresses to a level 
of $10^{-9}$ all Fourier terms with periods smaller than 66.7 
yr (stop-band), and attenuates to a level of $10^{-9}$ all Fourier terms
with periods between 66.7 and 200 yr. Terms 
with periods larger than 200 yrs are in the passband (see added information on website).  
By following this procedure, we have eliminated all 
frequencies connected with the orbital periods of Saturn and the satellites.

Finally, the shape of the frequency peaks can be modeled as Gaussian, but 
our measurements are made at isolated points.  This introduces what in signal 
processing is called ``impulse noise'', which can be attenuated by the use 
of a median filter (Press {\em et al.} 1996, Pitas 2000).  More details about 
the use of the median filter and of the ${\chi}^2$ test we devised to prove
the efficacy of our filter can be found in the Appendix.

To find an alternate way to distinguish between regular and chaotic 
behavior, we also computed Maximum Lyapunov Exponents (MLE) for our set 
of initial conditions.  A detailed explanation of the theory of Lyapunov 
exponents goes beyond the scope of this paper; instead we 
refer the reader to Lyapunov (1907) and Benettin {\em et al.} (1980).  
MLE is a measure of exponential stretching of nearby orbits.
The Lyapunov exponents are equal to zero 
for regular orbits (they tend to zero in finite-time calculations), while they 
assume positive values for chaotic orbits.  The inverse of a Lyapunov exponent
is the Lyapunov time $T_L$.  Smaller values of $T_L$ indicate enhanced 
local stochasticity.

To estimate MLE for orbits we 
used a modified version of SWIFT-LYAP2, a code that 
integrates the difference equation (Morbidelli 2002) 
in the SWIFT package (Levison and Duncan
1994).  The code was modified to reduce the integration error by integrating  
the planets in the heliocentric frame and the satellites in 
the planetocentric frame. For each of the test particles, we 
integrated the difference equation with an 
initial difference vector of modulus 
$d(0)=\sqrt{6}\cdot 10^{-9}$ and determined the modulus $d(t)$ of the 
displacement vector between the 
two vectors at time t.  We constructed a series $[t,\ln{[d(t)/d(0)]}$ and 
performed a linear least-squares fit on this series.  Since 
$d(t) \simeq d(0)\exp{(Lt)}$, 
where $L$ is the Lyapunov exponent, the slope of $\ln{[d(t)/d(0)]}$ versus
time is equal to the Maximum Lyapunov exponent. 
We computed the Lyapunov exponents for all orbits in 
our low-resolution survey.  We will
show these results in the following section.

We should point out that FAM and MLE, while both useful tools to investigate
chaotic behavior, measure different things.  FAM measures macroscopic changes 
in frequencies (i.e., the speed at which chaotic motion 
changes the frequencies), while $T_L$ measures the rate of exponential 
stretching of nearby trajectories.  These two techniques 
are complementary: 
FAM is more practical in finding regions that are macroscopically 
unstable, and $T_L$ can reveal the presence of different chaos-generating 
mechanisms.  We will discuss in more detail these aspects of the two methods 
in the following section.

\section{Origins of Chaos in the Kozai Resonance}

\begin{figure}
\plotone{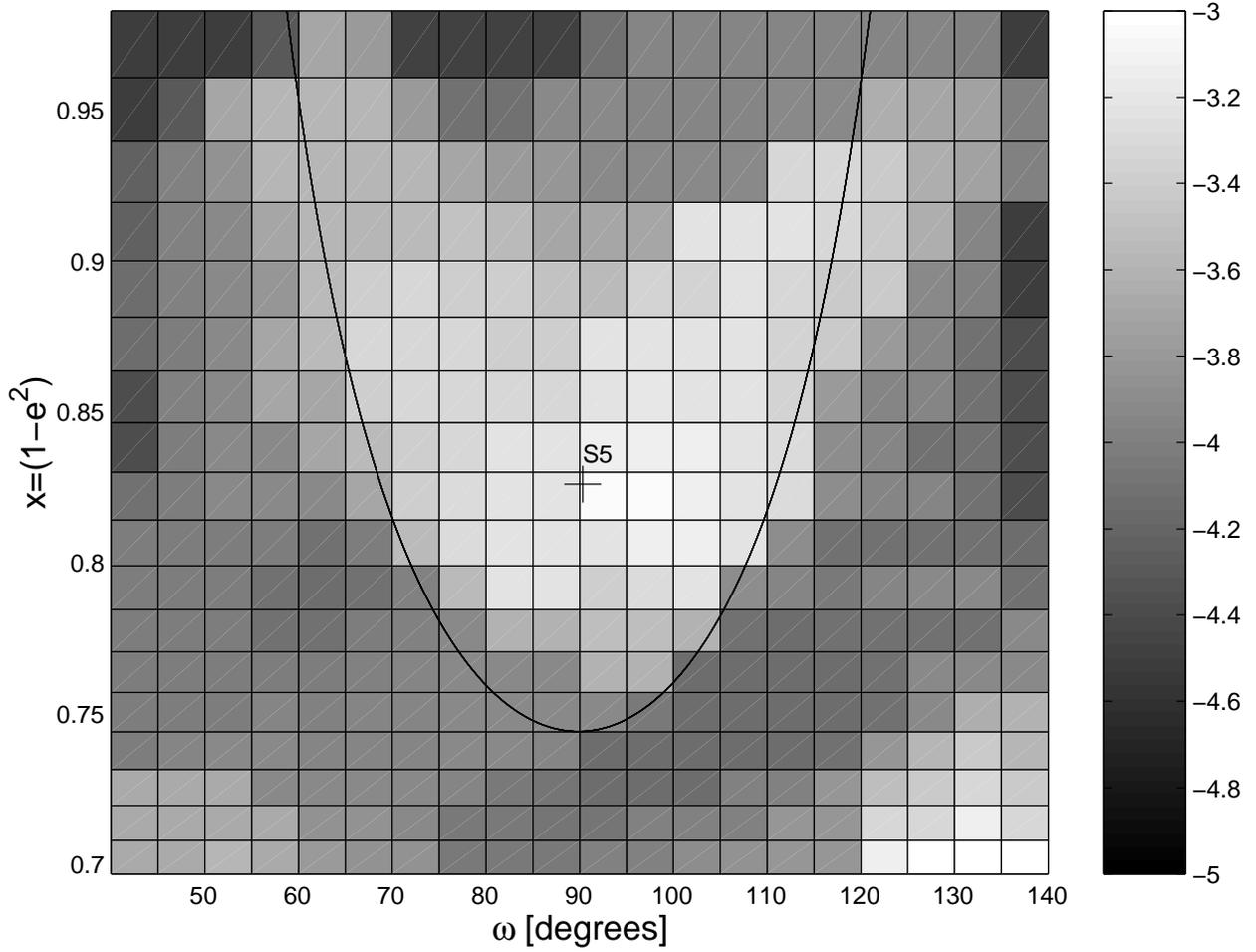}
\caption{Low-resolution survey, grayscale plots of 
$\sigma$ values for $\Delta T$ =1 Myr.  The 
solid parabolic arc in the figure's center represents
the separatrix according to the secular model (see Fig.~2).  
Chaotic orbits are associated 
with lighter shades (i.e., higher values of $log_{10}(\sigma)$, 
(see grayscale at right).
\label{fig: fig_lr1}}
\end{figure}

To investigate the presence of chaos, 
we applied the FAM to our low-resolution survey, integrated under the 
influence of the four jovian planets.
Fig.~\ref{fig: fig_lr1} plots $\log{\sigma}$ 
for our low-resolution 
survey when the time interval between the first and the second frequencies was 
about 1 Myr.  Fig.~\ref{fig: fig_lr1} shows that high values of 
$\sigma$ are found near the actual separatrix.
The chaotic layer is asymmetric between values of $\omega > 90^{\circ}$ versus
$  \omega < 90^{\circ}$, and 
Kiviuq seems to be on the border of the chaotic layer.

Having shown that a chaotic layer exists, the next logical 
step is to ascertain its origin(s).
Let us recall the analytical
model presented in Section 2.  In that model, the shape of 
the separatrix depends on Saturn's semiminor axis $b'$, which is a function
of the semimajor axis and the eccentricity.  Variations in
Saturn's semimajor axis and eccentricity may therefore generate 
complex orbital evolution near the separatrix.
But on what time-scales?  And how can we prove this hypothesis?

To study whether a variable $b'$ may generate chaos, we performed
a numerical simulation with 
the Sun (Saturn) on an orbit of fixed eccentricity (equal to the 
average value of the saturnian eccentricity) and the set of test 
particles used for our low-resolution survey.  The difference compared 
to the integrable model is the presence
of short-period terms.  We 
performed the FAM on test orbits in our low-resolution survey and found that all orbits are regular in this case.  In contrast, when only 
Saturn and the Sun were present, and Saturn 
had a constant eccentricity equal to the mean value
computed during the 46,000 yr period of oscillation (= $1/g_6$), 
no chaos was observed.  This seems 
to confirm our intuition that variations in Saturn's eccentricity 
and semimajor axis may be 
responsible for the chaotic layer.

\begin{figure}
\plotone{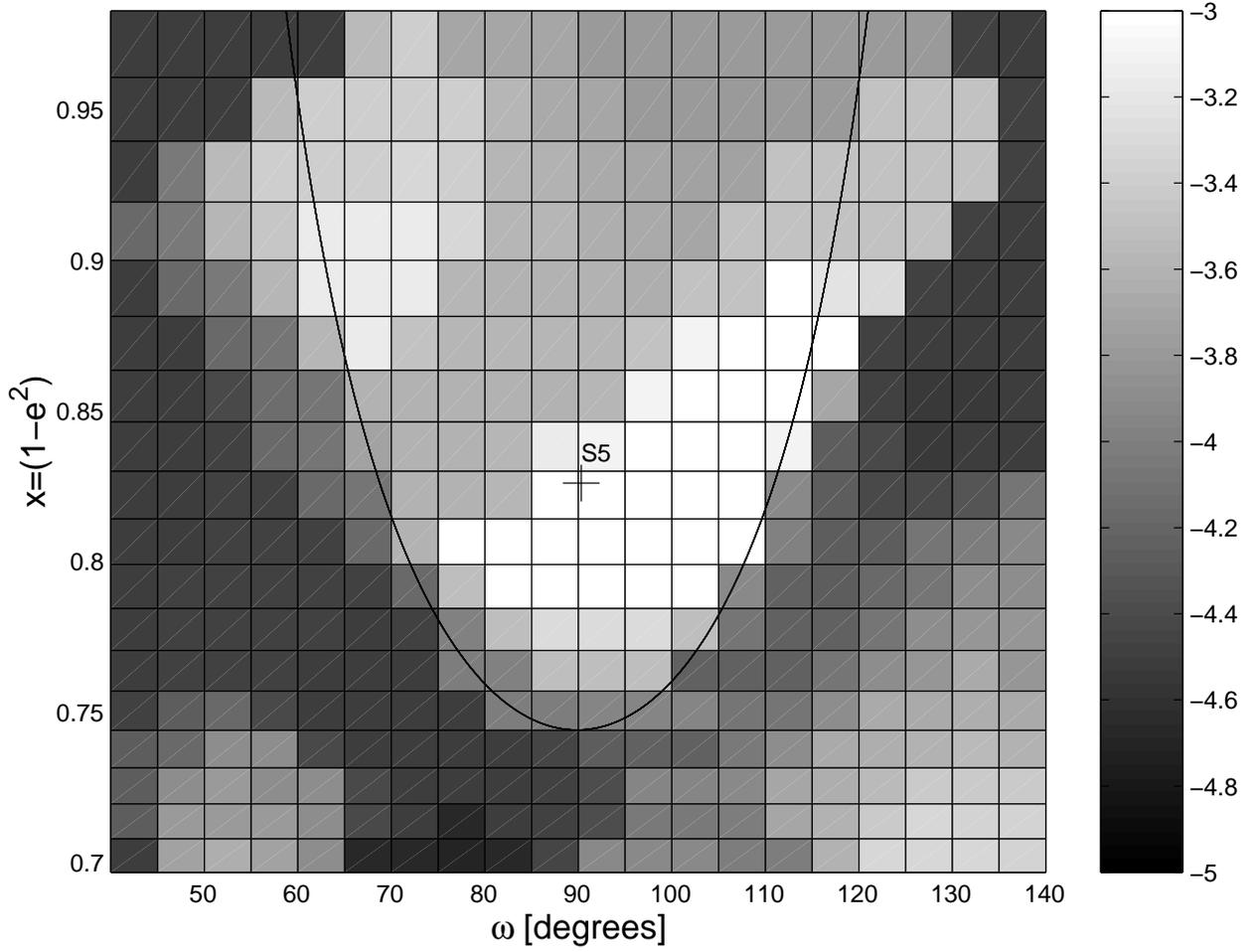}
\caption{The level of chaos for a simulation 
where particles are subjected to the gravitational 
perturbations from Saturn and Jupiter.  Chaotic orbits are in white, 
while orbits of regular behavior are in black (see grayscale at right). 
\label{fig: Saturn alone}}
\end{figure}

To test this hypothesis, we performed a simulation in which both 
Saturn and Jupiter are present.  In this case 
the layer of chaos appears (Fig.~\ref{fig: Saturn alone}).  
The similarity between Figs.~\ref{fig: fig_lr1} 
and ~\ref{fig: Saturn alone} suggest that the effects of 
Uranus and Neptune are unimportant.
To demonstrate this, we performed two simulations
with our low-resolution survey, 
one with the complete outer solar system (OSS), and the other with 
just Jupiter and Saturn.  
To show that the results are nearly equal we applied the
 ${\chi}^2$ test (see Appendix) to the filtered results of both 
simulations.  We obtained a value of ${\chi}^2$ of 22.4 (out of a 
maximum number of 399) indicating a very high 
probability that the two distributions are the same.
While perturbations from Uranus and Neptune might play a 
role in modifying the chaotic layer, such 
perturbations are clearly less important than Jovian perturbations.  
So, in the following, we only consider jovian perturbations.  In doing 
so we of course modify the values of planetary 
frequencies and mean motions, and therefore the positions of any 
secondary resonances. But, since in this phase of our work we 
are interested in understanding on what timescales the chaotic behavior is 
driven, we believe this is an acceptable approximation.

There are two prominent timescales for oscillations of Saturn's $e$ and 
$a$: the secular oscillations (period of order $1/g_6 \simeq 46,000$ yrs) 
and the oscillations related to the Great Inequality (883 yrs).
To study which of these perturbations is more relevant to the origin of chaos, 
we describe the OSS' orbital evolution, in terms of Fourier 
series (Bretagnon and Francou 1988).
In this model, the equinoctial orbital elements ($a,\lambda,K=e\cos{\varpi},
H=e\sin{\varpi},Q=\sin{\frac{I}{2}}\cos{\Omega},
P=\sin{\frac{I}{2}}\sin{\Omega}$) are 
developed in series of cosines and sines of combinations of twelve angles: 
four mean longitudes of the giant planets, four 
proper longitudes of pericenter, and four proper 
longitudes of the nodes. These proper angles are linear 
functions of time.  This 
model assures a fractional precision better than $5\cdot10^{-5}$ 
for $K,H,Q,$ and P over 10 Myr.

To investigate what minimal model can create a chaotic 
layer similar to the one observed in the full problem, 
we used different solutions 
for the orbits of both Jupiter and Saturn.  We 
modified our version of SWIFT-WHM so that both orbital evolutions 
of Jupiter and Saturn 
are computed ``on-line'' according to the Bretagnon secular 
solution.  The test particles are then subjected to perturbations of 
planets that evolve on such orbits.
We considered four models for Saturn's orbital evolution: (1) SEC, in  
which we limit the expansion to terms containing the $g_5, g_6,$ 
and $s_6$ frequencies for $K, H, Q,$ and $P$, assume 
a constant semimajor axis, and take 12 terms for $\lambda$'s development.  
(2) A21SEC, in which we have included 
12 terms associated with the 1:2 commensurability between Jupiter's 
and Saturn's
mean motions, and terms associated with the Great Inequality for the 
development in {\em a}, plus the same terms as in the SEC solution.
(3) GISEC, in which there were 8 terms for $K$ and $H$, and 6 for 
$P$ and $Q$ associated with 
the Great Inequality, and a constant semimajor axis; and (4)  
A21GISEC, which includes the 2:1 terms in {\em a}, the 
Great-Inequality terms and the secular variations.  Table~\ref{table: bret_model} gives the number of terms taken in $a, \lambda, (K,H), (Q,P)$ 
for each solution.

\begin{figure}

\plottwo{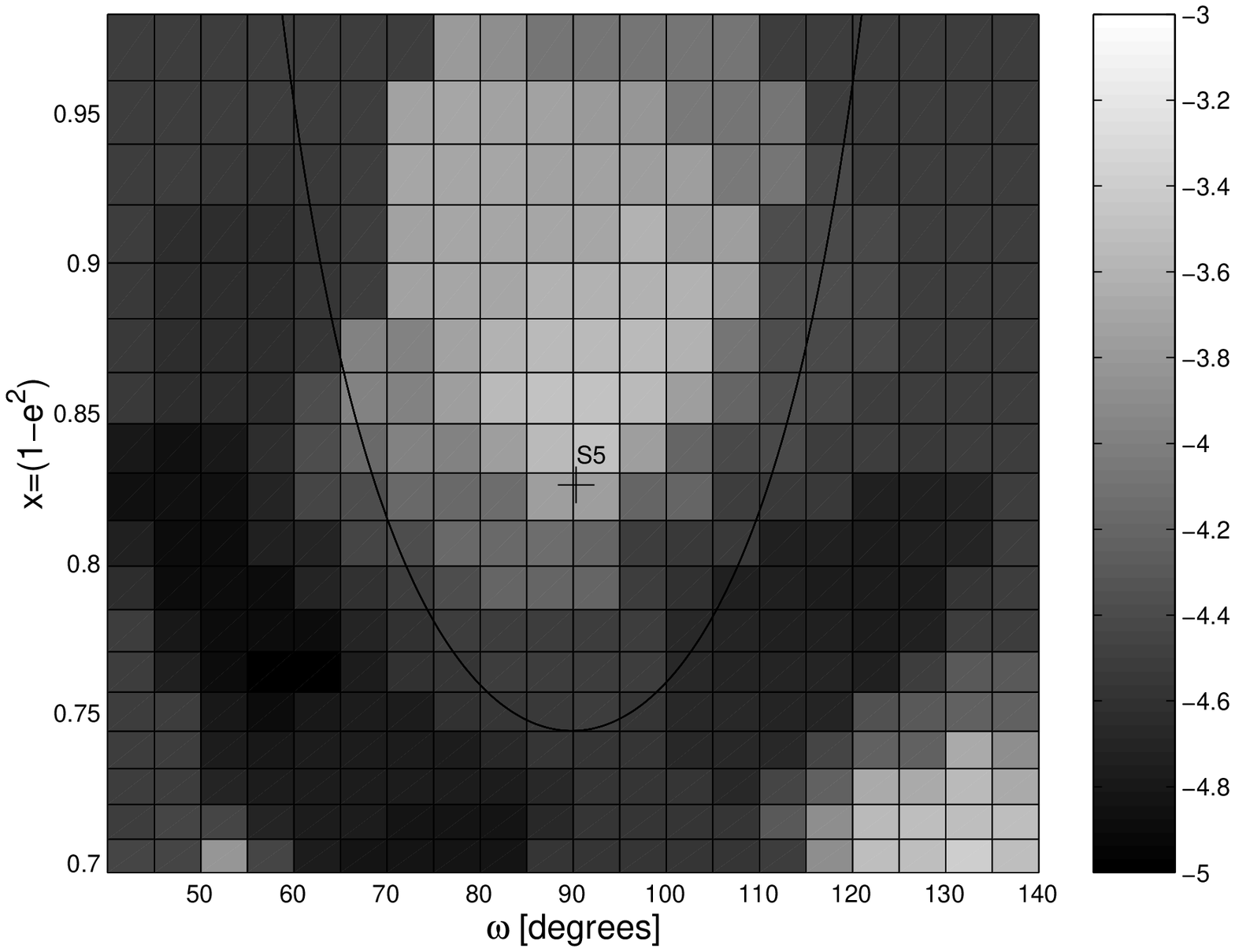}{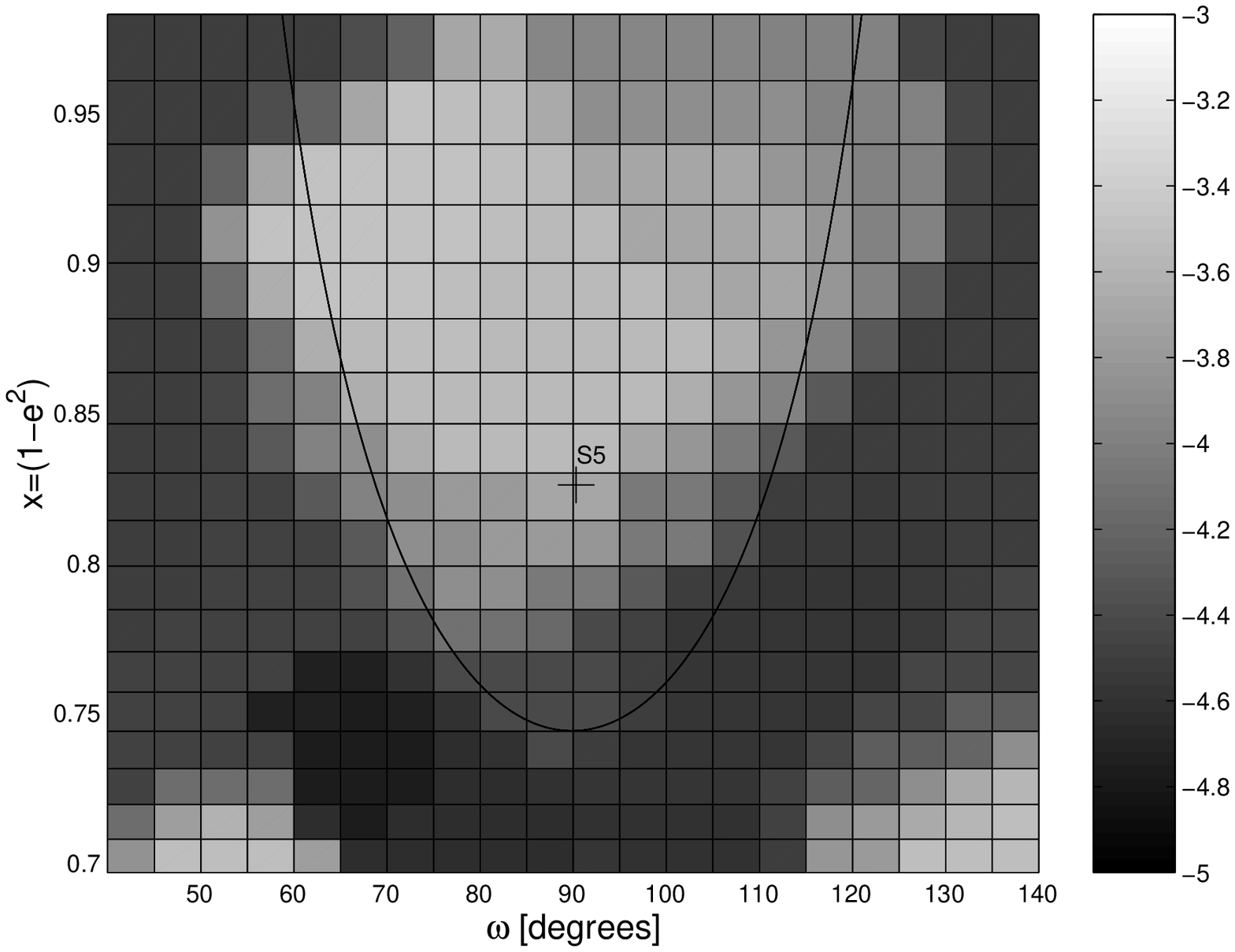}

\plottwo{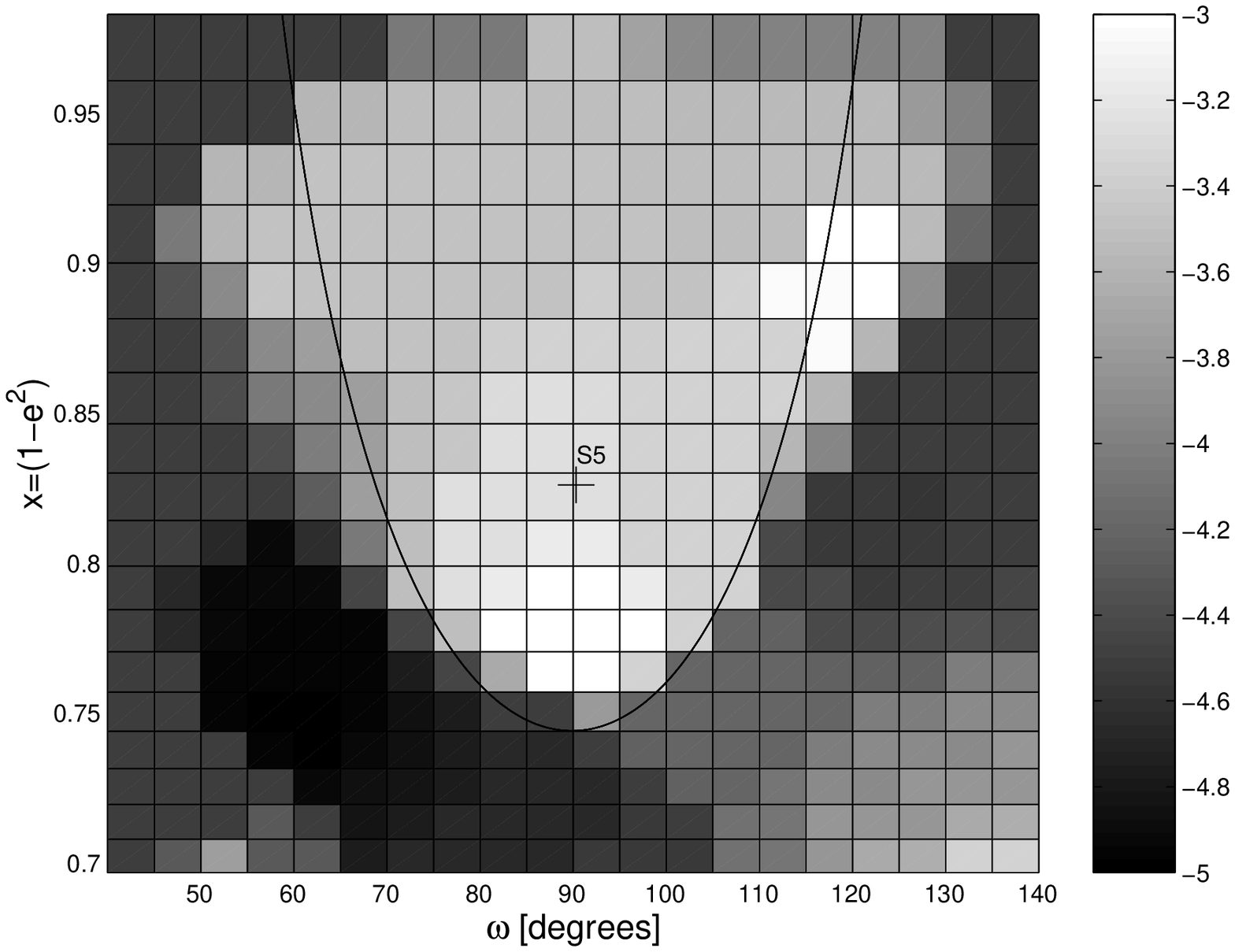}{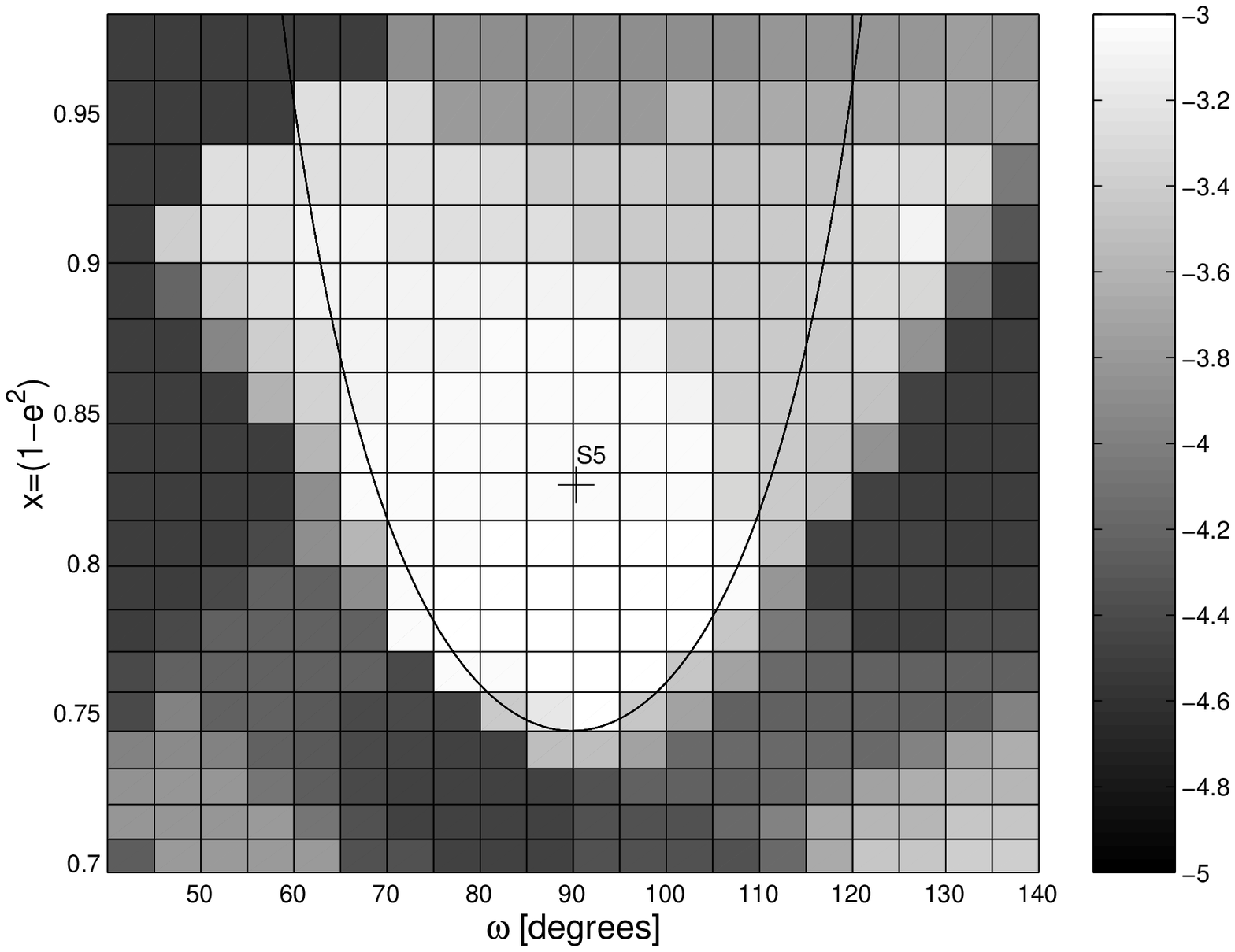}

\caption{Measures of chaos for the simulation with the SEC Bretagnon 
solution (upper left), for the A21SEC Bretagnon solution (upper right), 
the GISEC solution (lower left), and the A21GISEC solution 
(lower right).  Light shades correspond 
to $log_{10}({\sigma}) = -3$ or to more chaotic orbits.} 
\label{fig: SEC}
\end{figure}

Fig.~\ref{fig: SEC} shows the results of the integration with 
these models. All simulations used the same solution for 
Jupiter (10 terms in the development of $a$, 
2 terms for $K$ and $H$, 2 for $Q$ and $P$, and 
12 for $\lambda$).  These figures may be compared with 
Fig.~\ref{fig: Saturn alone}, where the traditional version of SWIFT-WHM 
was used.

Table~\ref{table: bret_res} presents the results of a ${\chi}^2$ test 
in which we compared simulations with ``Bretagnon'' models 
with those of the full integration of Jupiter and Saturn.  
According to these results, terms associated 
with the Great Inequality and the 2:1 terms in the development of 
Saturn's semimajor axis are the most important for the 
presence of the chaotic layer.  The major sources for chaos 
are the variations in Saturn's
eccentricity connected with the Great Inequality.

\begin{figure}
\plottwo{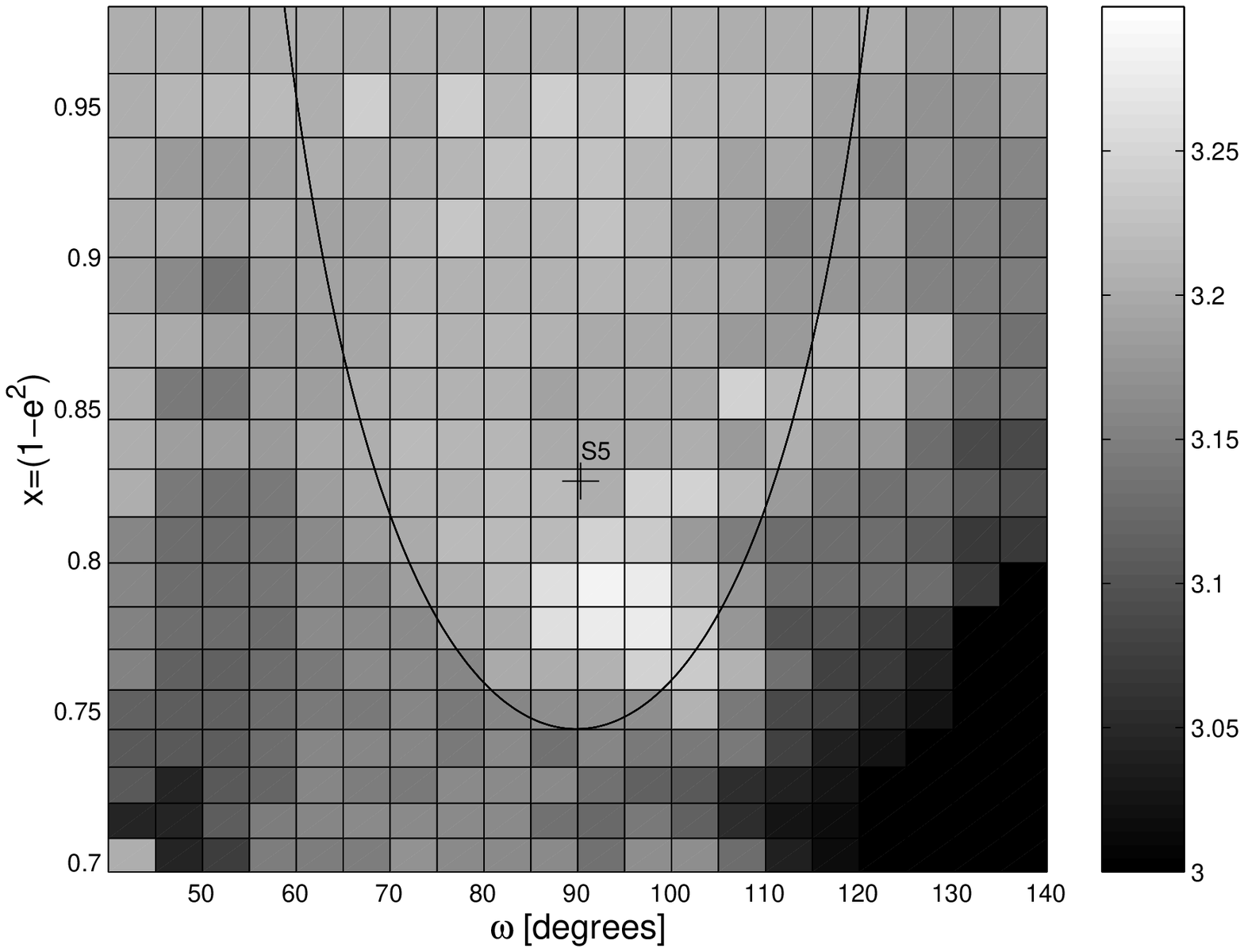}{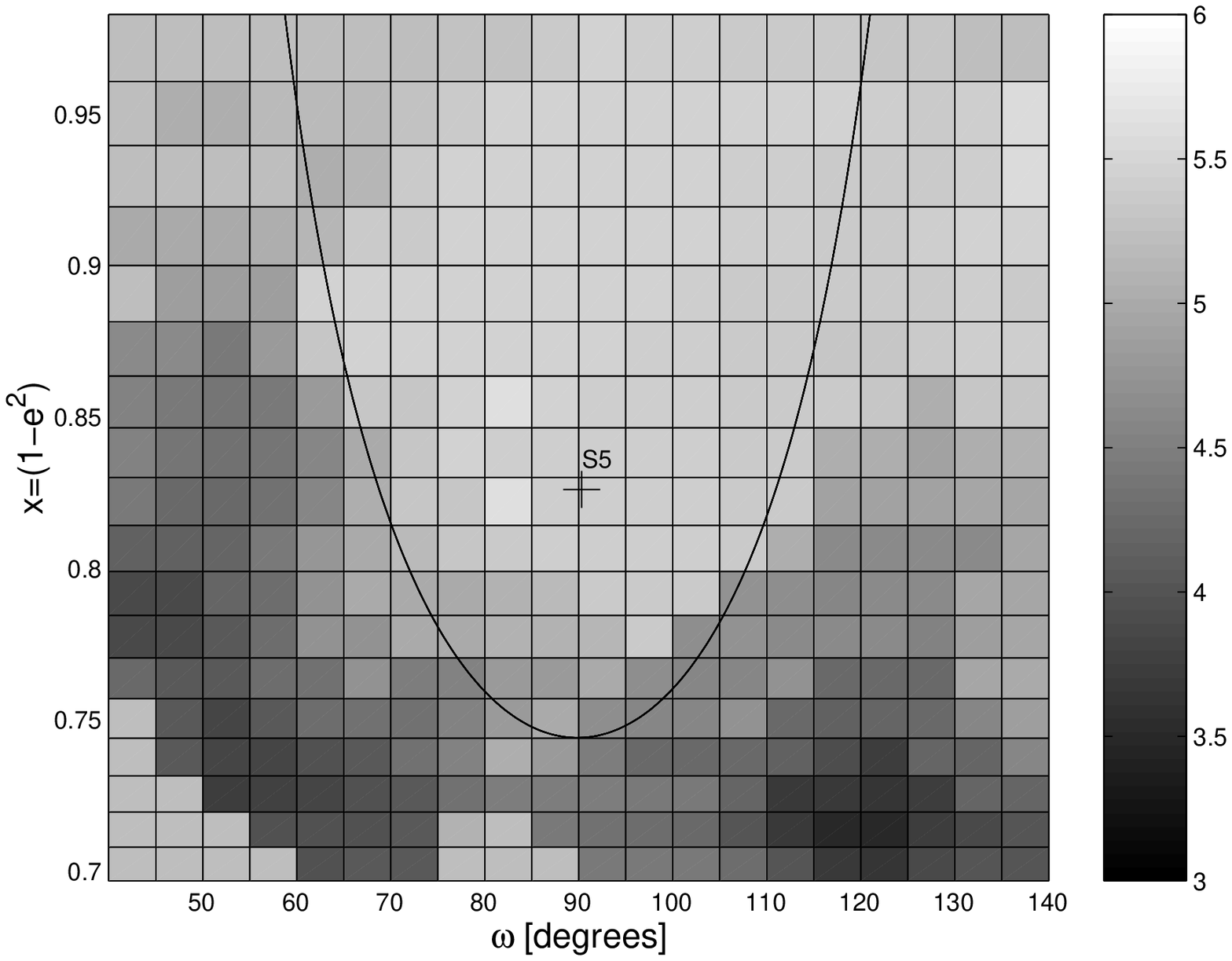}
\caption{Lyapunov time for integrations (left) when the traditional 
version of the Lyapunov integrator is used and (right) when the SEC 
solution of the Bretagnon model is applied.  The left figure has a feature 
of long Lyapunov times, associated with a region of high 
frequency dispersion, but on the right figure we see that this 
feature disappears when the SEC solution is used b).  The 
substantially different gray scales used in the two figures 
were chosen to maximize the contrast between nearby features of $T_L$.} 
\label{fig: Lyap_time}
\end{figure}

To further investigate this hypothesis, we computed the maximum Lyapunov
exponent for the particles of our low-resolution survey, in a model
containing Jupiter and Saturn. Fig.~\ref{fig: Lyap_time} 
shows the resulting values of $T_L$
(left) when the full problem is considered and (right) when the planets
evolve according to the SEC solution. A remarkable asset of 
Fig.~\ref{fig: Lyap_time}a is
the region of high values of $T_L$ (i.e., a smaller degree of local
stochasticity) near the vertex of the separatrix, which, according to the
FAM, is a region of high dispersion of the frequencies. In this region
$T_L\approx 1,700$ yrs, i.e., twice the period of the Great Inequality; 
thus our results suggest that the feature is connected to a 
secondary resonance with the Great Inequality.  Note, however, 
that all orbits seem to have quite small values of $T_L$.
Orbits inside the libration island, which, according to FAM, are
macroscopically regular, have $T_L<2,000$ yrs. This appears to be a general
characteristic of high-inclination irregular satellite orbits, as
all seem to have $T_L<10,000$ yrs (M. Holman, private communication,
2003).

The existence of a region characterized by relatively long Lyapunov times,
on one hand, but high frequency dispersion, on the other, seems
counter-intuitive at first. However, recall that the
rate of frequency dispersion is a macroscopic characteristic of
the orbit that is related to the global structure of the phase space region.  
In contrast, the Lyapunov time is a local property whose value is
dictated by the mechanism that is responsible for the chaotic motion. 

The existence of regions of large $T_L$ but rapid diffusion is not uncommon in
the asteroid belt. For example, in the 7/3 resonance (see Tsiganis 
{\em et al.}, 2003), the smaller values of $T_L$ are 
observed at the borders of the
resonance where they are related to the pulsation of the 
separatrix due to secular
precession. However, the macroscopically most unstable region is
located at smaller libration amplitudes, is characterized by larger values
of $T_L$, and is related to the colocation of the $\nu_6$ secular
resonance. Similar results can be shown for other low-order mean-motion
resonances (e.g., 2:1 and 3:1), where the most unstable zones are generated
by coexisting secular resonances. As $T_L$'s value is typically of
the order of the forcing period (e.g., $1/g_5\approx 305,000$ yrs,
$1/g_6\approx 46,000$ yrs, Tsiganis and Morbidelli 2004), these regions
generally have higher values of $T_L$ than the region associated with the
pulsating separatrix of the mean-motion resonance.

\begin{figure}
\plotone{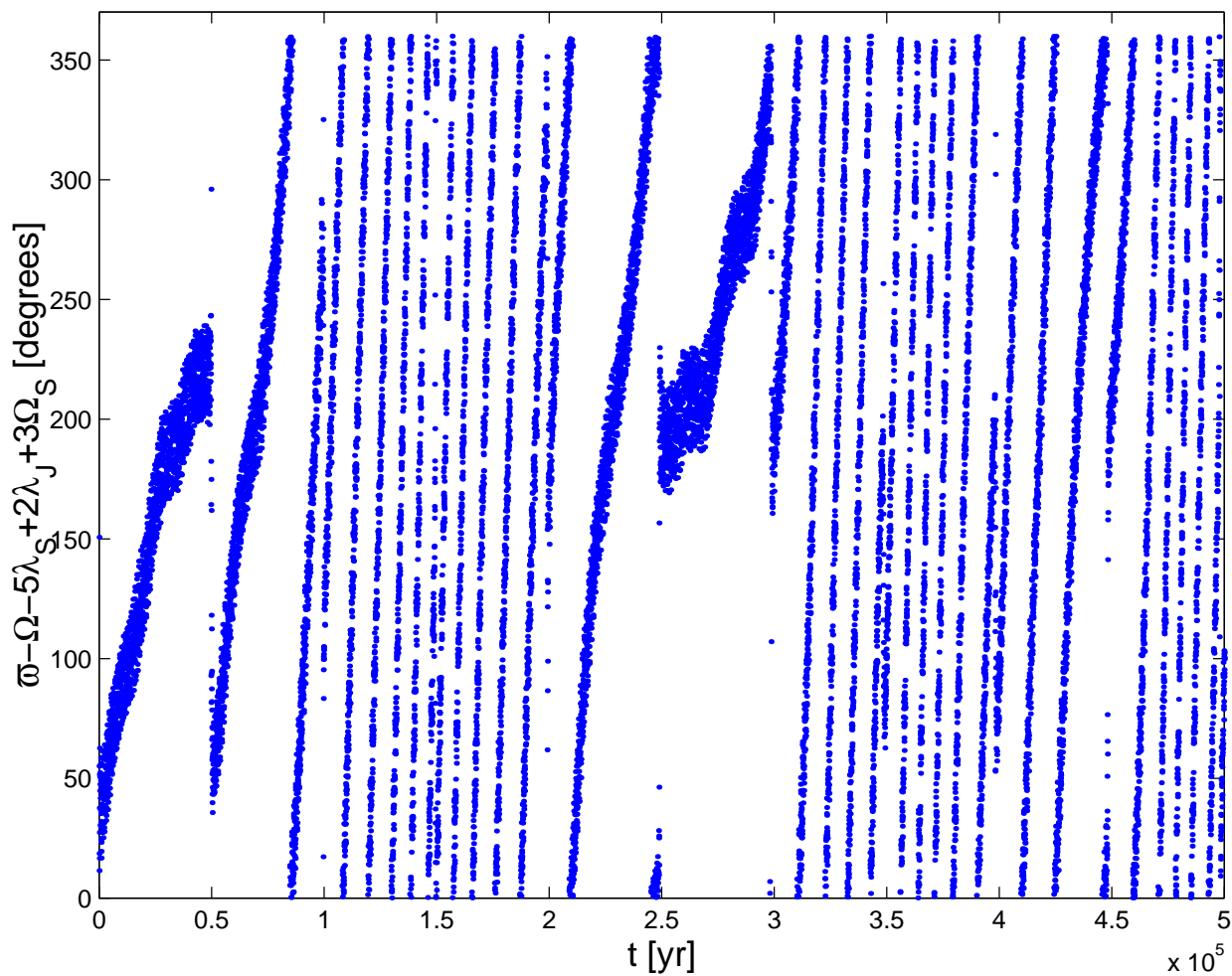}
\caption{Time evolution of the angle $\varpi-\Omega-
5{\lambda}_{S}+2{\lambda}_{J}+3{\Omega}_{S}$ 
for an orbit in the region of the secondary resonance involving the 
Great Inequality ($x_0 = 0.756, {\omega}_0 = 90^{\circ}$).  Note how 
the resonant argument alternates between 
periods of libration and periods of circulation.  This particle
has a large value of $\sigma$.
\label{fig: resonant_arguments}}
\end{figure}

We are likely observing a similar phenomenon in the case studied here.  
We believe that this high-$T_L$ region is
associated with a secondary resonance, involving 
the Great-Inequality period. This is also supported by the results shown in 
Fig.~\ref{fig: Lyap_time}b (SEC solution), where all 
orbits inside the libration zone become regular
(note the different scale compared to Fig.~\ref{fig: Lyap_time}a). 
To further support our
claim, we selected a particle from this high-$T_L$ region, for which we
plot the time evolution of the argument $\varpi -\Omega
-5\lambda_S+2\lambda_J+3 \Omega_S$ (Fig.~\ref{fig: resonant_arguments}). 
This corresponds to one of the
possible critical arguments of the 1:1 resonance between the frequency
of $\omega=\varpi-\Omega$ and the frequency of the Great Inequality,
$2\lambda_J-5\lambda_S$ ($3\Omega_S$ is added in order to fulfill the d'
Alembert rules of permissible arguments, i.e., the sum of
coefficients of individual longitudes must be zero and the sum of coefficients
of nodal longitudes be even). The behavior of the arguments switches
erratically between intervals of libration and circulation, which
indicates a transition through the separatrix of this secondary resonance.

The following picture arises from our low-resolution surveys.
Two major sources of chaos exist in the system.
One is connected to the transition region from circulation to libration and, 
in particular, is driven by variations in Saturn's eccentricity, mainly 
associated with Great-Inequality terms, and variations in Saturn's 
semimajor axis.
The other is related to a secondary resonance involving the 
argument of pericenter $\omega$ and the Great-Inequality terms 
for circulating particles.  In most cases this resonance is 
a stronger source of chaos than the transition region.  Better resolution is 
needed to understand the behavior and shape of this resonance (see Sec. 5).

One problem that remains to be explained 
is the asymmetry of the chaotic layer with respect to $\omega=90^{\circ}$ 
(see Figs. 5-7).  
According to the secular
model, the result should be symmetric.
Our simulations with the Bretagnon model suggest that variations in 
Saturn's semimajor axis may be partly responsible for the asymmetry.
Alternatively, the asymmetry in the chaotic layer 
may be related to our choice of initial conditions.  To exclude 
this second possibility, we need to 
ascertain that our choice of initial conditions ($\Omega$ and $M$) 
for the test particles' orbits is 
not introducing spurious effects.   To our knowledge, 
three resonant configurations may  
alter the eccentricities of the test particles, and so 
put their orbits closer or farther away from the region of chaos that we are 
interested in.  The three resonances are: "nodal" 
($\Omega-{\Omega}_{\odot} = $constant), pericentric 
($\varpi-{\varpi}_{\odot} = 0$), 
and the evection inequality [$2(\varpi-{\lambda}_{\odot})=$constant 
($= 45^{\circ}$, so as to have an intermediate value of the evection 
angle)\footnote{Since the evection term in our case is not resonant 
(Nesvorn\'{y} {\em et al.} 2003) we refer to the angle 
$2(\varpi-{\lambda}_{\odot})$ as evection inequality.}.    In our 
simulation, the Sun's orbit starts with: $\Omega=23.162^{\circ}, 
\omega=311.992^{\circ}$, and  $M=322.291^{\circ}$.  
We computed $\Omega$ for the 
test orbits according to the resonant configuration.  For the pericentric
resonance, we used two values of the resonant argument: 
$\varpi-{\varpi}_{\odot}=0$ and  
(Yokoyama {\em et al.}, 2003), $\varpi-{\varpi}_{\odot}=180^{\circ}$, 
$\lambda-\lambda_{\odot}=180^{\circ}$.
The second configuration maximizes the perturbing effect of the pericentric
resonance.  

To have a quantitative measure of the asymmetry, we use 
an ``asymmetry coefficient'' that is the difference of $\sigma$'s values for 
$\omega$ larger and smaller than $90^{\circ}$.  To compute this coefficient
for orbits in the libration island ($60^{\circ}<\omega<120^{\circ}$),
we take two columns symmetric about $\omega=90^{\circ}$ 
(e.g., the columns for $\omega=60^{\circ}$ and $\omega=120^{\circ}$, 
with the obvious exclusion of the column for $\omega=90^{\circ}$) and 
compute the fractional difference for each pair.  We repeat the 
process for all pairs of values and compute 
the average and standard deviation of the measurements.  The average 
value is our ``asymmetry coefficient'', and the standard deviation gives 
an estimate of the error. Obviously, the lower is the asymmetry coefficient, 
the more symmetric is the distribution of $\sigma$ around 
$\omega = 90^{\circ}$.

Table~\ref{table: asy_res} gives
asymmetry coefficients for each simulation.
Since the secular resonance with resonant argument equal to 
$0^{\circ}$ seems to give the more symmetric configuration, 
we believe its effect might be the dominant one (but large errors prevent our 
firm conclusion).  

\section{A Web of Secondary Resonance Crosses the Kozai Resonance}

The low-resolution survey has yielded information regarding 
the large-scale structure of chaos inside the Kozai resonance.  
However, because most of the chaotic behavior is present near the boundary
region between circulation and libration, we now concentrate on 
analyzing this region in more detail.
For this purpose, we constructed a new set of initial conditions: we used 
276 bins in inclination, starting at $37.5^{\circ}$ and separated by 
$0.02^{\circ}$.  We used one $\omega$ (= $90.0^{\circ}$).  
The choice of the other
orbital elements followed the same criteria used for the low-resolution 
survey.  We called this new set of initial conditions a ``high-resolution''
survey.  We concentrate on the region around $\omega = 90^{\circ}$
to eliminate the problems associated with asymmetries in 
$\omega$ suggested by our low-resolution survey. Results of integrations
of test particles with $\omega$ equal to $89.8^{\circ}$ or $90.2^{\circ}$
are very similar to those with $\omega = 90^{\circ}$ and are not shown.
Our choice of values for the particles' inclinations 
allow us to sample periods in libration and circulation that go from a minimum 
of 450 yrs to a maximum of 820 yrs for librating particles, and from a minimum
of 650 yrs to a maximum of 1820 yrs for circulating particles, 
thus covering the whole transition region.

\begin{figure}
\plotone{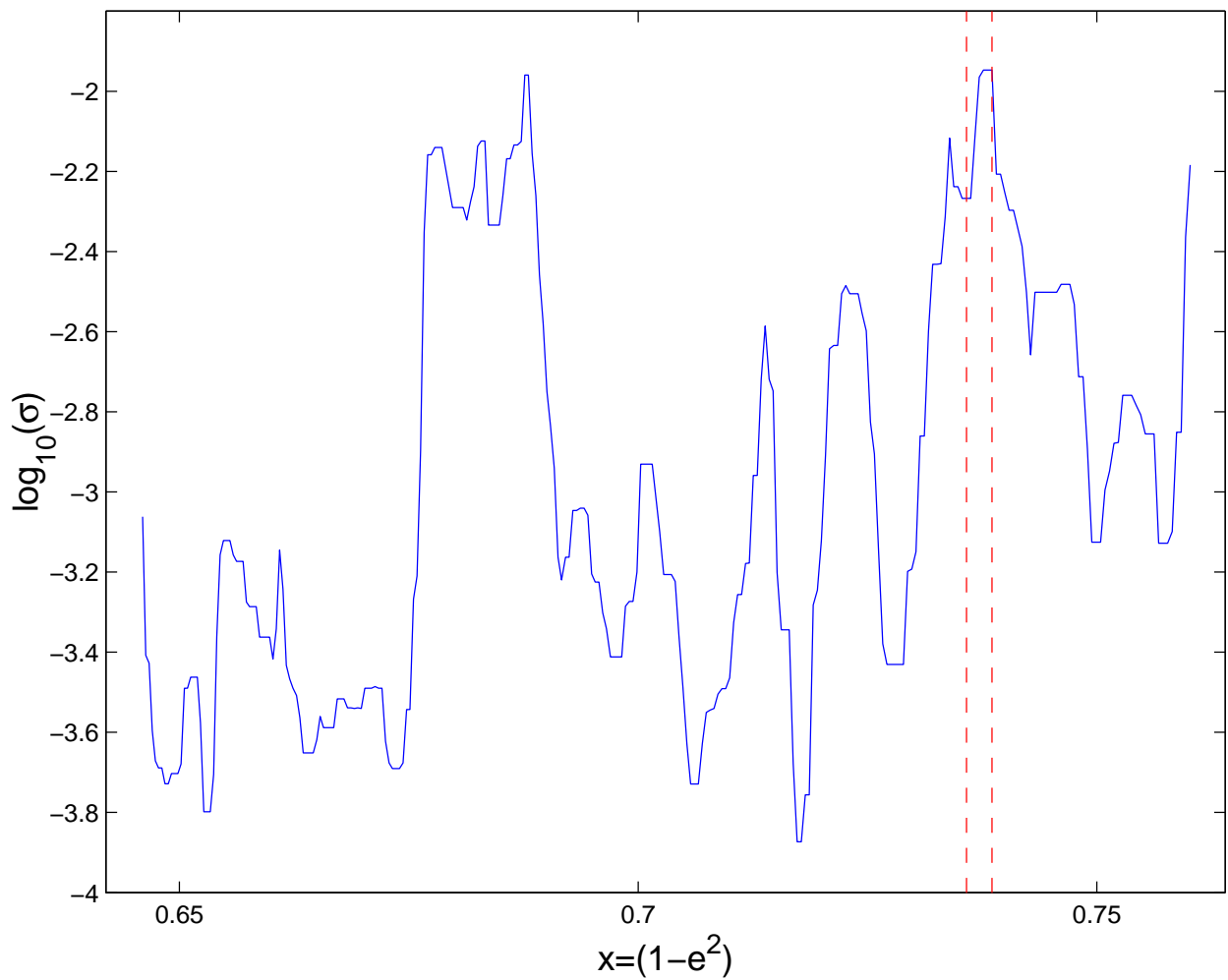}
\caption{A plot of $log_{10}(\sigma)$ versus initial $x$ 
for our high-resolution survey.
The dashed vertical lines show the transition region in which 
particles may switch behavior from circulation to libration. The two 
major sources of chaos for the system are this transition region and 
the pericentric secular resonance at $x$ = 0.68.
Especially important in the context of planetary migration is the 
secondary resonance between $\omega$ and the Great Inequality at $x$ = 0.696.  
Other secondary resonances are also present.
\label{fig: high_res_survey}}
\end{figure}

\begin{figure}

\plottwo{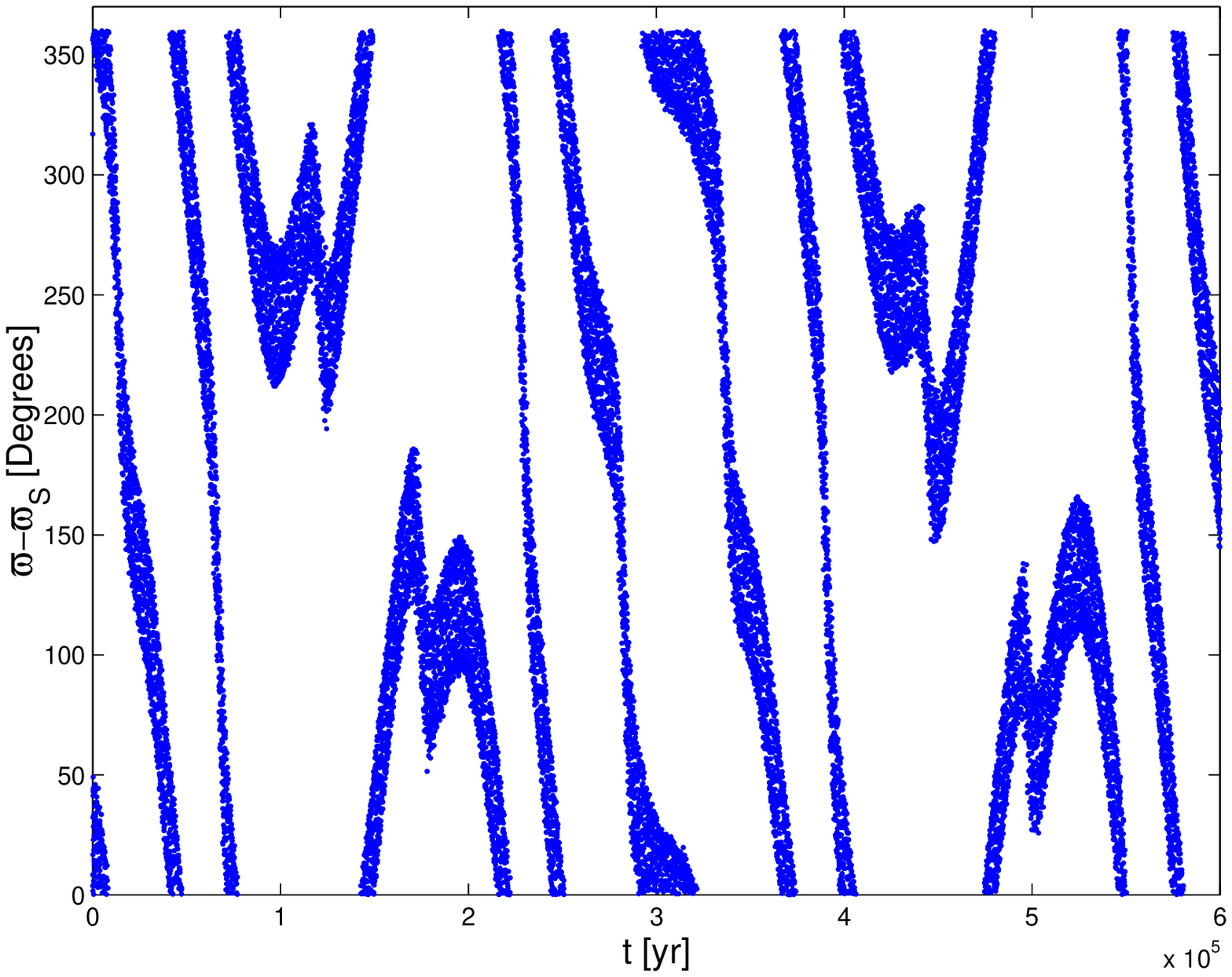}{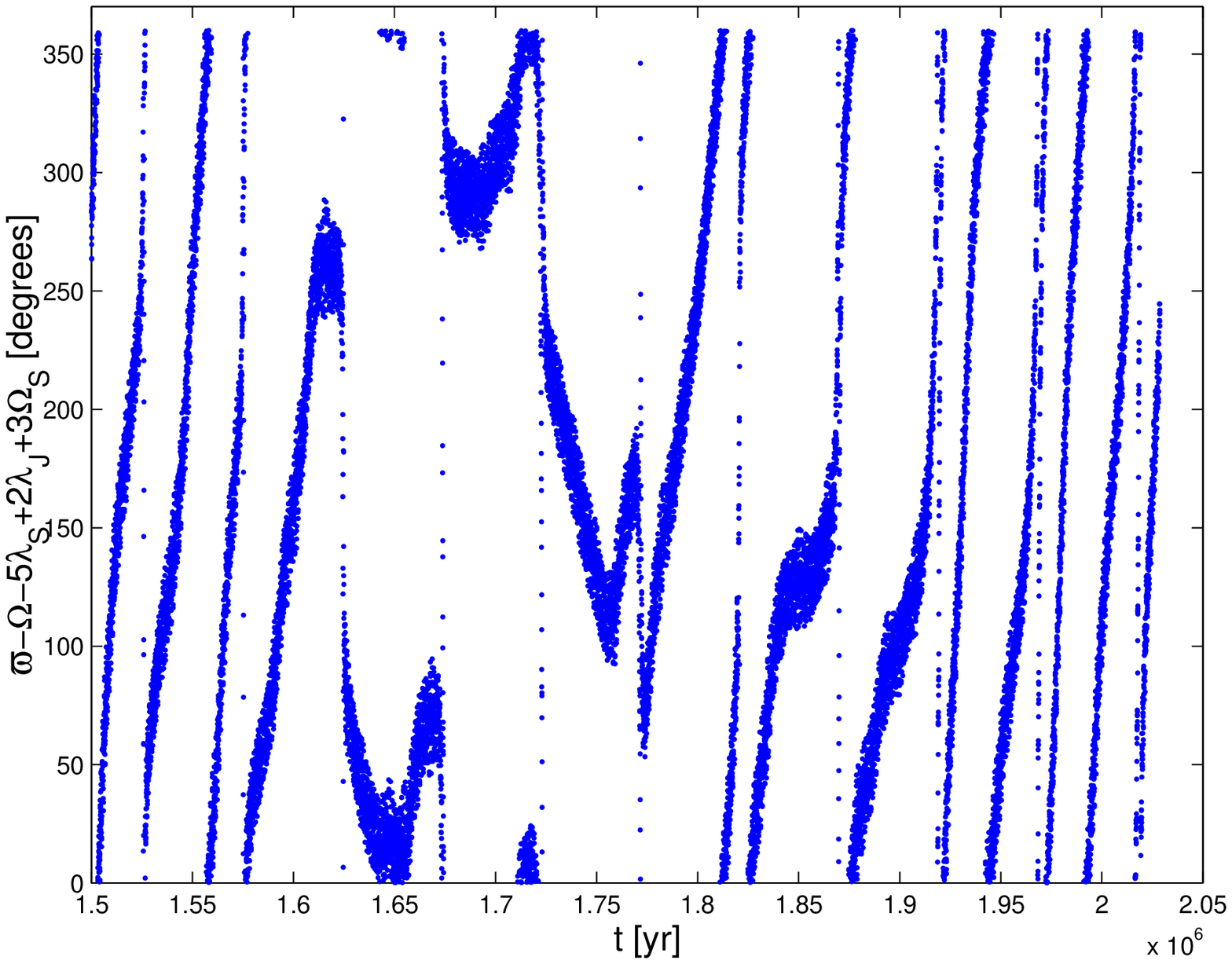}

\plottwo{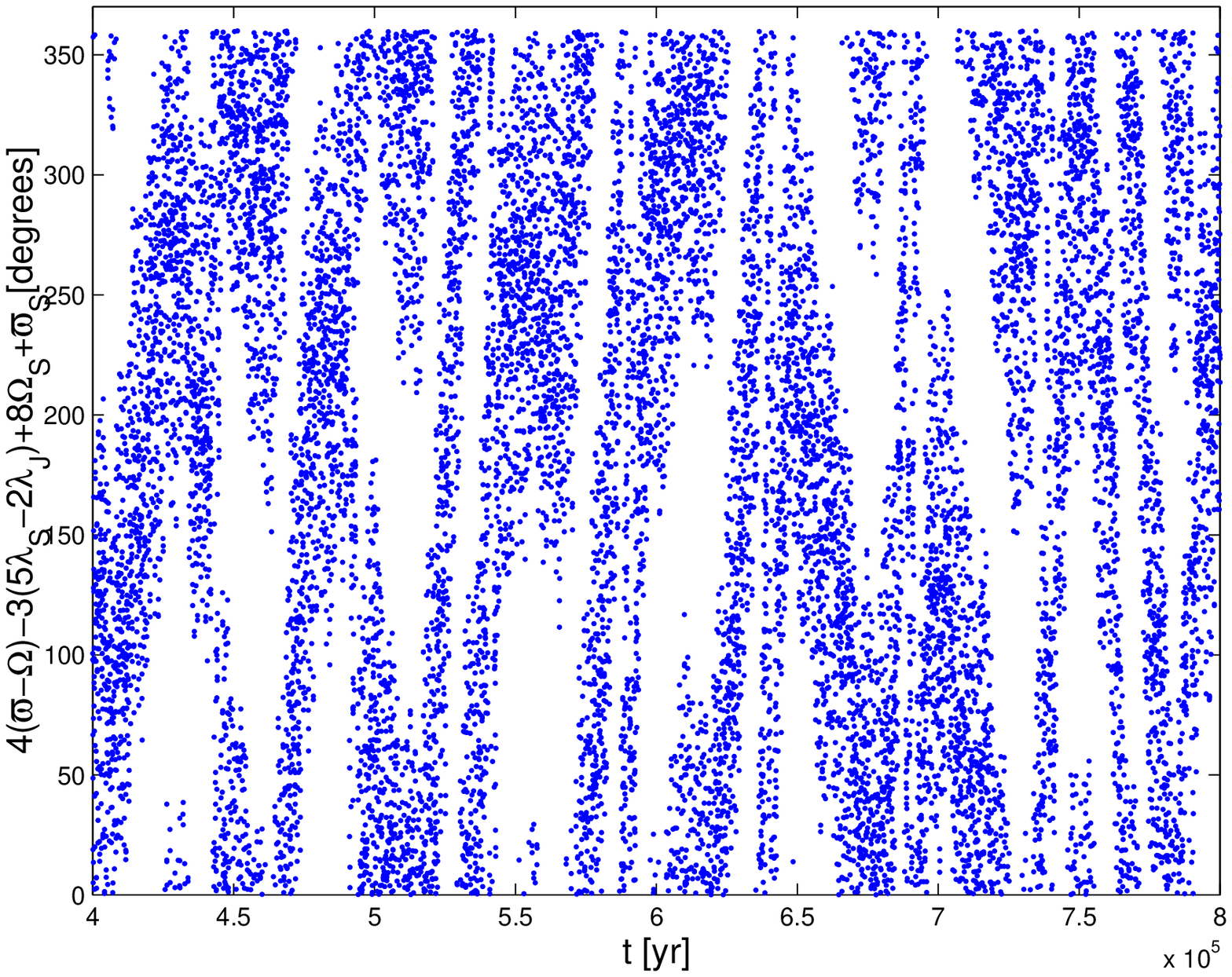}{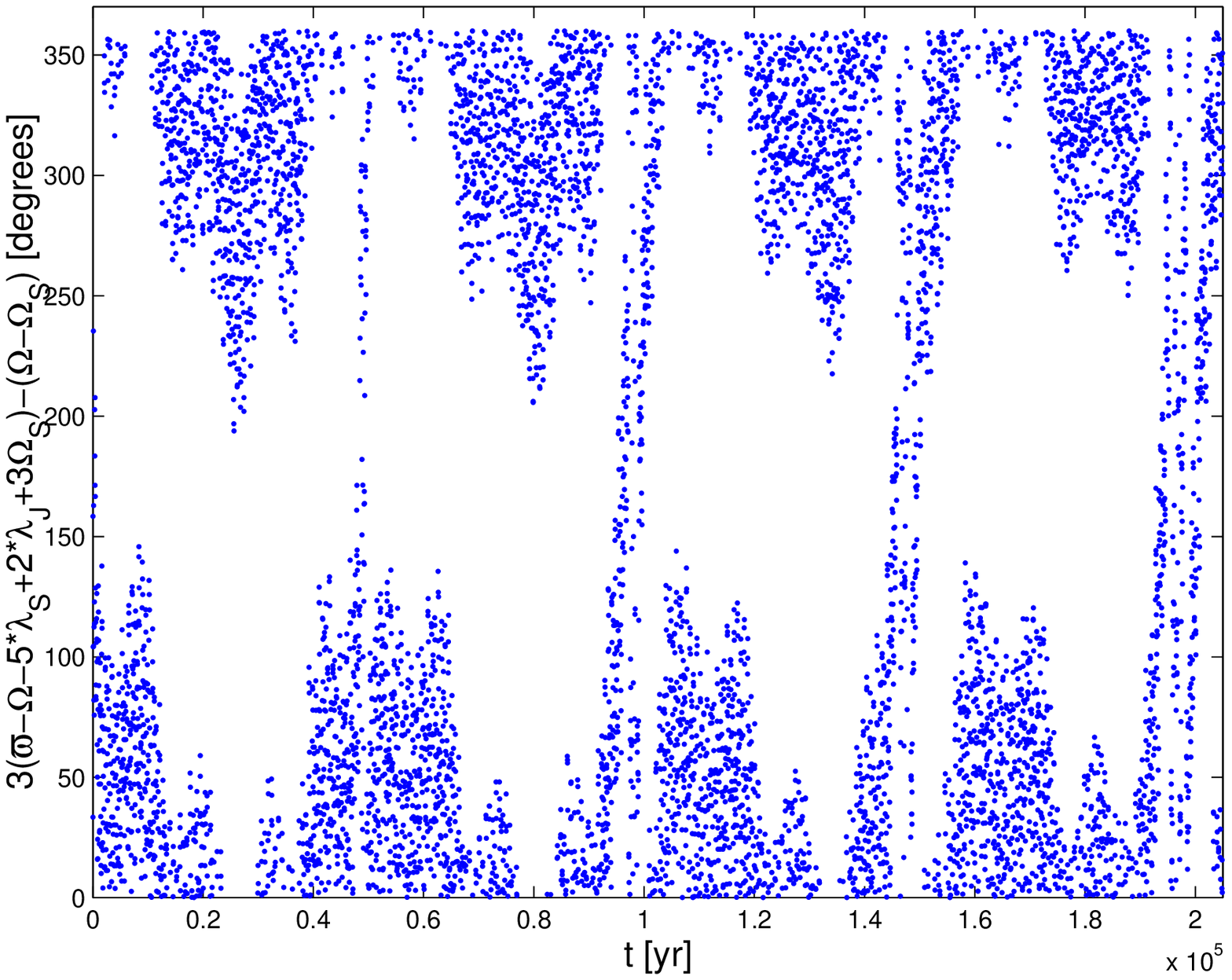}

\caption{Resonant arguments for test particles in the a) pericentric
secular resonance, b) resonance of argument 
$\varpi-\Omega-5\lambda_{S}+2\lambda_{J}+3\Omega_{S}$, c) resonance
of argument $4(\varpi-\Omega)-3(5\lambda_S-2\lambda_J)+8\Omega_S+\varpi_S$, 
and d) $3(\varpi-\Omega-5\lambda_S+2\lambda_J+3\Omega_S)+(\Omega-\Omega_S)$.
All these particles shows periods of libration for their respective 
resonant angles.  The orbital elements of planets and satellites are 
computed with respect to the solar system's invariable plane.}
\label{fig: Res. Arguments}
\end{figure}

We integrated the test particles of our high-resolution survey under the 
influence of the four jovian planets for 2 Myr and applied FAM.  
Fig.~\ref{fig: high_res_survey} shows two regions of high chaoticity.
The one at $x \simeq 0.73$ is clearly associated with the 
transition from circulation to 
libration: test 
particles switch from $\omega$-circulation to $\omega$-libration 
at this $x$ (dots 
in the plot).
The other region of strong chaotic behavior 
at $x \simeq 0.68$ is associated with the 
pericentric secular resonance ($\varpi-{\varpi}_{S}=0$).   
Fig.~\ref{fig: Res. Arguments}a
shows the time evolution of the resonant argument for an orbit in this resonance ($x_0 = 0.6848$).  

\begin{figure}
\plotone{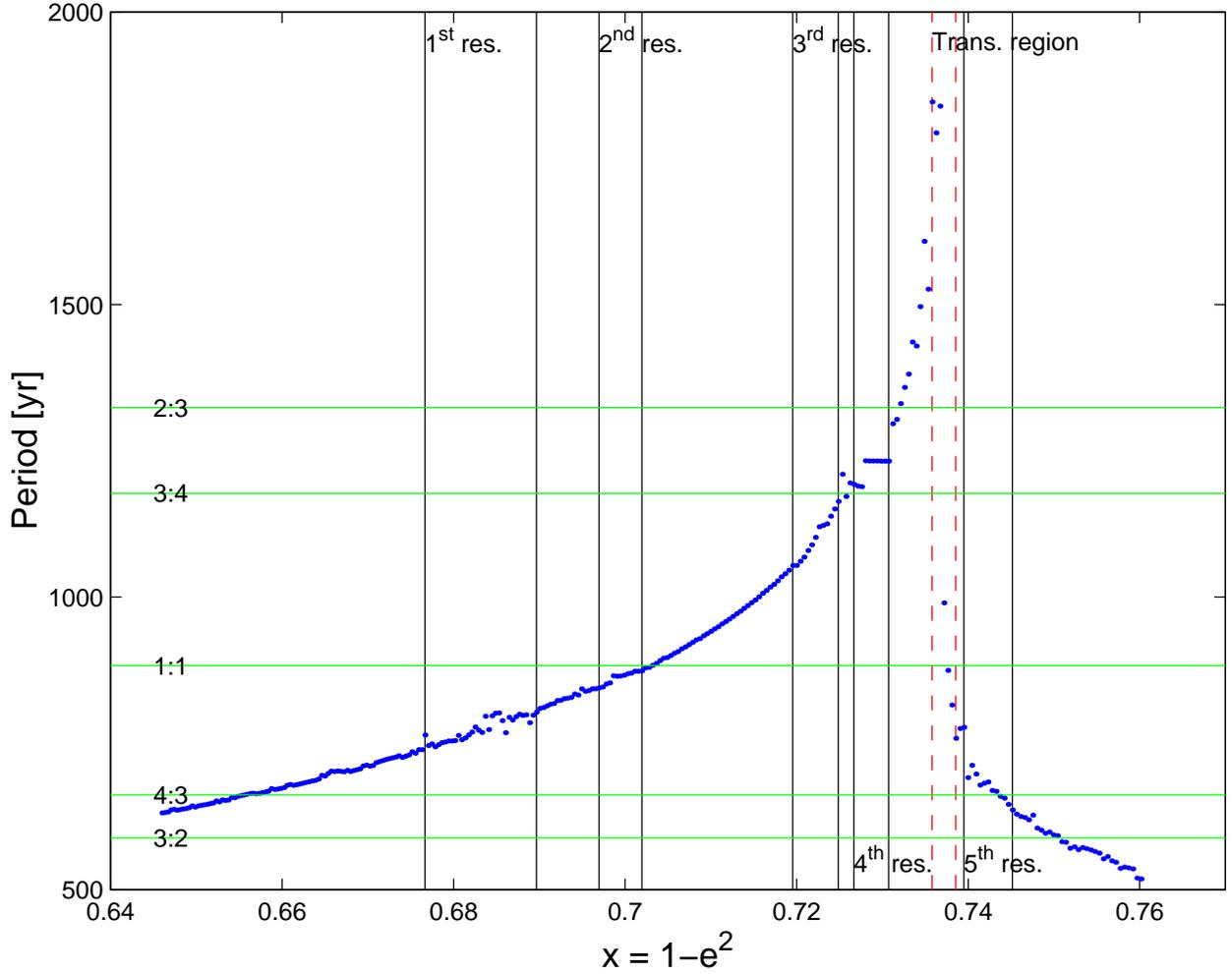}
\caption{Periods of $\omega$'s precession obtained with FAM as a function 
of $x$.  Vertical lines represent the locations of features of high 
$\sigma$, the dashed vertical lines show the transition region in which 
particles may switch behavior from circulation to libration, and 
the horizontal lines report the location of the commensurabilities 
between $\omega$ and the Great Inequality.  The positions of the actual 
resonances slightly differ from that of the commensurabilities (see
Eq.~\ref{eq: f_resonant}).
\label{fig: omega Periods}}
\end{figure}

More interesting for our purposes is the region at $x \simeq 0.696$.
A plot of the resonant argument $\varpi-\Omega-5\lambda_{S}+2\lambda_{J}+
3\Omega_{S}$ shows that this feature is associated with the 1:1 resonance 
between $\omega$ and the Great Inequality (other resonances involving 
$\varpi_S$ and $\varpi_J$, instead of $\Omega_S$, are also present
but are weaker.  This property is generally shared by other resonances involving 
harmonics of the Great-Inequality period, like the 4:3, 3:2, etc.).  
The resonance we discovered is only present for 
circulating particles.  A question that might arise is if a similar resonance 
could also be found for librating test particles.   To answer this 
question we consider the periods of libration and circulation and estimate them by the frequency 
associated with $\omega$'s precession found via FAM.  
Fig.~\ref{fig: omega Periods}
shows such periods as a function of $x$.   We also report the regions having 
high values of $\sigma$, possibly associated with secondary resonances 
(vertical lines), and the commensurabilities between periods in $\omega$
and Great-Inequality periods (for example, the 3:4 commensurability means 
that, for three periods in $\omega$ there are four periods in 
$2\lambda_J-5\lambda_S)$.  

We are conscious that the 
commensurabilities between the Great Inequality and $\omega$ are 
not real resonances, since they do not respect the d' Alembert rules.  
Nevertheless, we think this is a useful plot, since gives a 
first-order estimate of the position of the real resonances.
To compute the expected positions of the actual resonances 
we used the following 
procedure: for a resonance of argument $\varpi-\Omega-5\lambda_S+2
\lambda_J+3\Omega_S$ (we will use $\omega$ for $\varpi-\Omega$ hereafter), 
the frequency of $\omega$'s precession for which there is resonant 
behavior is given by:

\begin{equation}
f_{\omega}=f_{GI}-3f_{s_6},
\label{eq: f_resonant}
\end{equation}

\noindent where $f_{GI}$ is the frequency associated with the 
Great Inequality term
($2\lambda_J-5\lambda_S$, equal to 1467.72 ``/yr for the current position
of the planets), $f_{s_6}$ is the frequency associated with $\Omega_S$
(= -26.345''/yr, Bretagnon and Francou 1988, the same source is used 
for the values of $g_5$ and $g_6$), and 
$f_{\omega}$ is the precession frequency.
Analogous equations can be used for other resonances.  From $\omega$'s
frequency it is straightforward to determine the period
of precession, and from that period we can determine the value of $x$.
Fig.~\ref{fig: omega Periods} seems to be very instructive.
For example, since the period of the 1:1 resonance is 840 yrs and 
since the first 
test particle in librating regime has a period of 820 yrs, at the end 
of the transition region, there is currently 
no equivalent of the Great-Inequality secondary resonance for librating 
particles.  
Also, apart from the two regions with $\sigma$ associated 
with the pericentric secular resonance and the Great-Inequality resonance,
another interesting feature of high chaotic behavior, which seems to be 
associated with a strong secondary resonance, appears at $x=0.73$.  
Fig.~\ref{fig: omega Periods} 
shows how in this region the values of periods in $\omega$ are 
constant, which is an indication of librating behavior in a secondary 
resonance.   Other regions of high $\sigma$ values can be found at 
$x = 0.722$ and, in the librating region, 
at $x = 0.741$ (since this region is so close
to the separatrix, the chaotic behavior here observed might be related
to the pulsating behavior of the separatrix itself, when perturbations 
from other jovian planets are considered).

We believe that the features of high $\sigma$ at $x = 0.722$ and $x = 0.73$ 
are associated with two other secondary resonances, whose resonant arguments are given by $4(\varpi-\Omega)-3(5\lambda_S-2\lambda_J)+8\Omega_S+\varpi_S$ and
$3(\varpi-\Omega-5\lambda_S+2\lambda_J+3\Omega_S)+(\Omega-\Omega_S)$, 
respectively.  Fig.~\ref{fig: Res. Arguments}c and d 
show the behavior of the argument for 
orbits near these two resonances.   Other weaker resonances connected 
with other commensurabilities between $\omega$ and $2\lambda_J-5\lambda_S$
are also observed.  The resonance of argument 
$3(\varpi-\Omega)-2(5\lambda_S-2\lambda_J)+5\Omega_S+\varpi_S$ 
is expected to be 
stronger than the one associated with the 3:4 commensurability, but its 
location is so close to the separatrix that its effect is difficult to 
discern.  Table~\ref{table: Resonant Arguments} lists the resonances for which
the librating behavior of the resonant argument is observed, and a few 
candidates that could explain features of weaker chaos.

So far we have only discussed the case of circulating test particles.  Features of weaker chaos, however, exist also in the librating region.  Unfortunately, in this case plotting the resonant argument is not so easy, since, by the definition of libration, $\omega$ oscillates around $\pm 90^{\circ}$ 
instead of covering all values from $0$ to $360^{\circ}$.
A possible way to overcome this problem would be to make a change of 
variables, 
so as to put the origin of the system of coordinates for $e\cos{\omega}$
and $e\sin{\omega}$ at the libration center.  The new angle $\omega'$ would 
then rotate from $0$ to $360^{\circ}$, and it would then be possible to 
check for the behavior of the resonant arguments by combining $\omega'$ 
with GI and other terms.  
This procedure is rather cumbersome, especially since the libration point 
itself is not a fixed point, but, due to perturbations from the other jovian 
planets, oscillates with 
timescales associated with the Great Inequality, $g_5, g_6, 
s_6$, etc.   Fortunately, there is an alternative method to determine whether
a resonance has in its resonant argument terms associated with the Great
Inequality.

\begin{figure}
\plotone{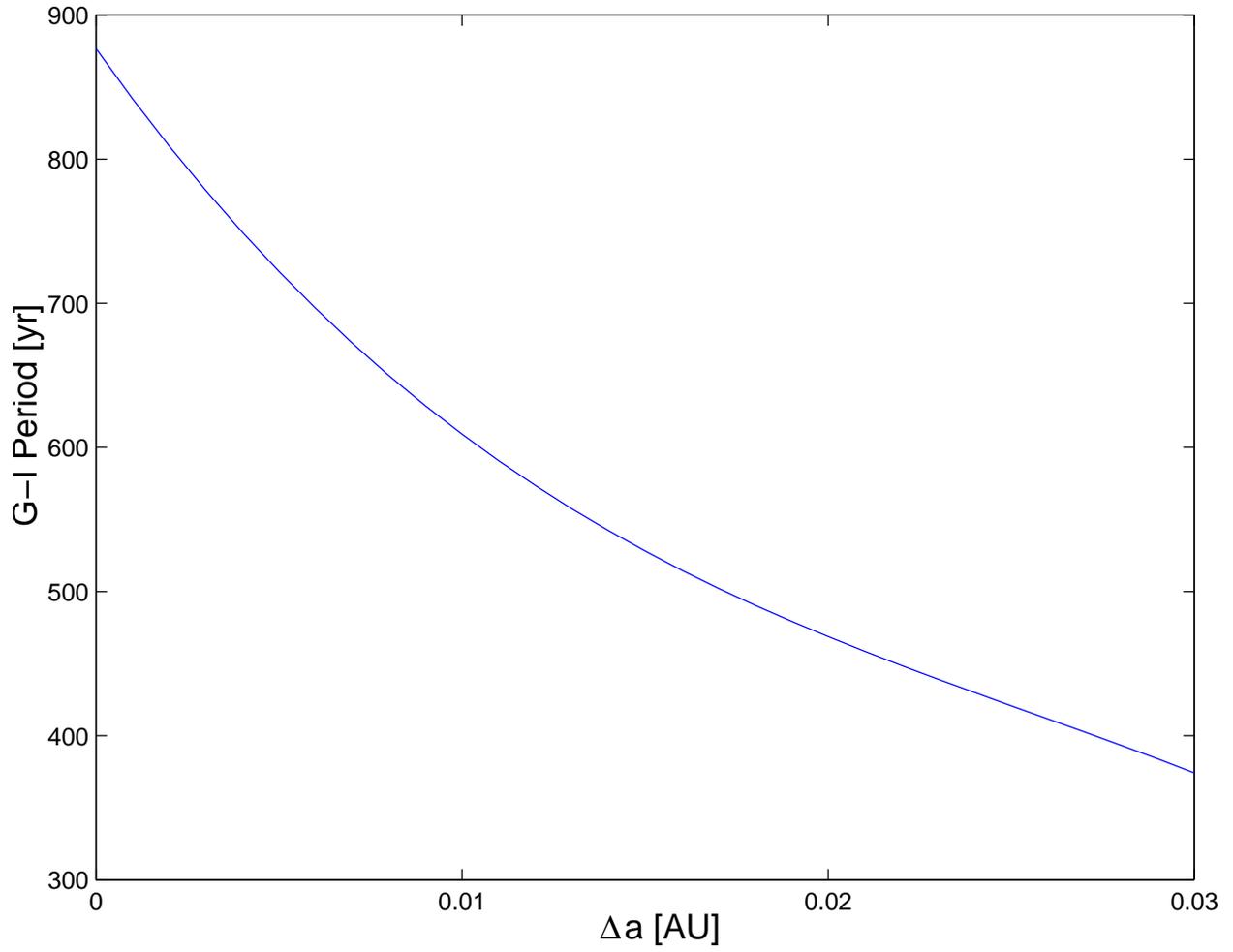}
\caption{The period of the Great Inequality as it varies with positive 
expansions in Jupiter semimajor axis, assuming Saturn's orbit fixed.
\label{fig: Great-Inequality period}}
\end{figure}

It is widely believed that planets have migrated since their formation
(Malhotra 1995, Gomes 2003, Levison and Morbidelli 2003).   In particular, 
the gravitational scattering of a 
planetesimal disk modified the positions of the jovian
planets so that, while 
Saturn, Uranus and Neptune migrated outward in semimajor 
axis, Jupiter, being the most effective scatterer, migrated inward.
As the work of Malhotra suggested, the origin of the highly eccentric, 
inclined, and Neptune-resonance-locked orbits of Pluto and the Plutinos might 
be explained in the context of sweeping resonant capture due to the 
changing position of Neptune's orbit.  Our work has shown that a major 
source of chaos for test particles in Kozai resonance around Saturn is 
due to a secondary resonance between the argument of pericenter $\omega$,
the argument of the Great-Inequality (i.e., 
$2{\lambda}_{J}-5{\lambda}_{S}$), and terms connected with the
planetary frequencies $g_5, g_6,$ and $s_6$.
It is well known that by altering the positions of Jupiter 
and Saturn, the period of the Great Inequality changes.  
Fig.~\ref{fig: Great-Inequality period} shows how, by varying the initial 
value of the osculating 
semimajor axis of Jupiter by a positive amount (and keeping 
the position of Saturn fixed) it is possible to modify the period
of the Great Inequality from the current value of 883 years to values 
of $\simeq 400$ yrs or less. \footnote{A 
similar change can also be achieved by 
modifying the initial value of the difference in mean longitude of 
Jupiter and Saturn (Ferraz-Mello {\em et al.} 1998).  
This alternative method has the advantage of better
preserving the values of the secular frequencies of the planets (i.e., 
$g_5, g_6, s_6$).  However, in this work, we are interested in the actual 
effect of planetary migration, which alters the semimajor axes of the 
planets, and so we prefer to use the first method.} 

\begin{figure}
\plottwo{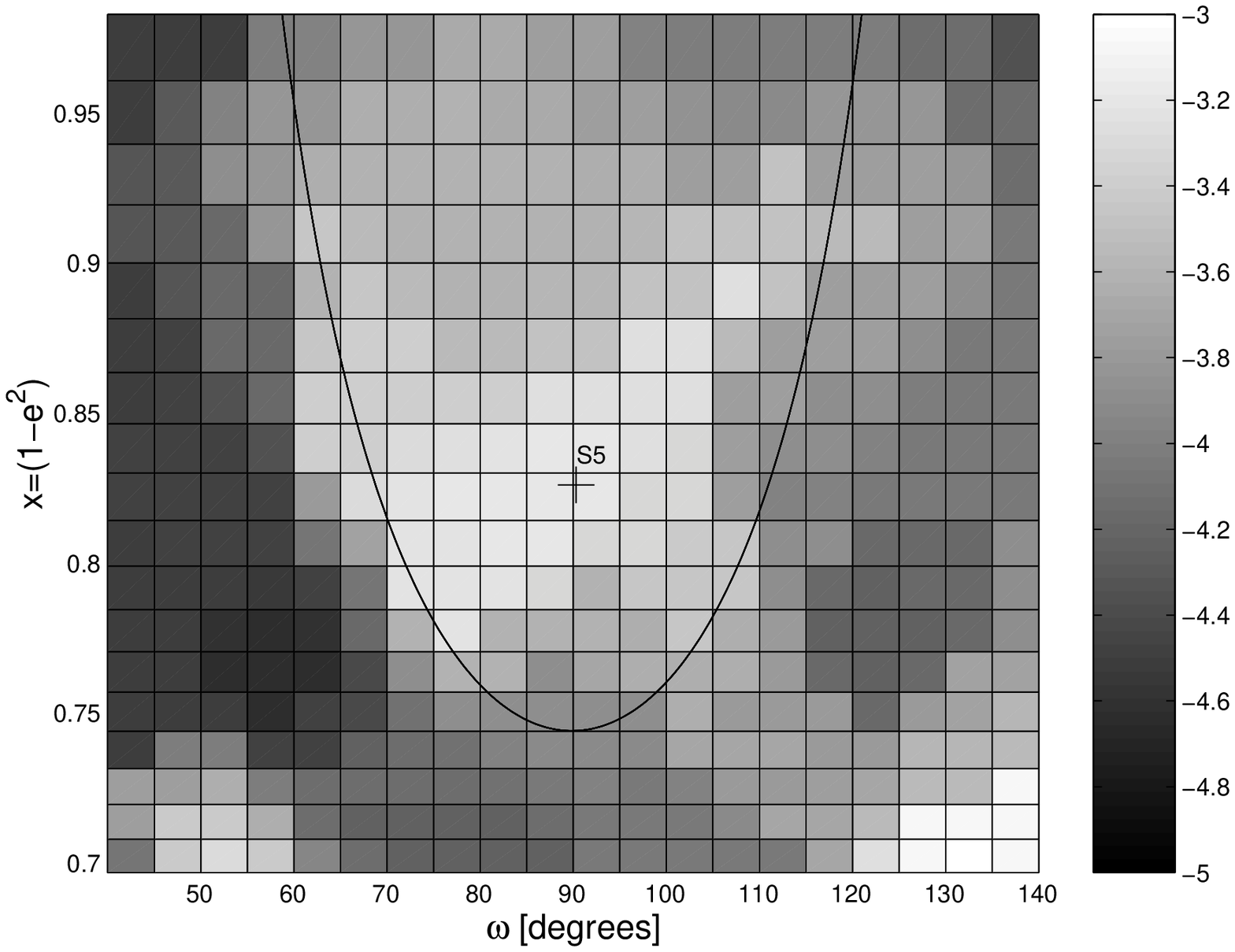}{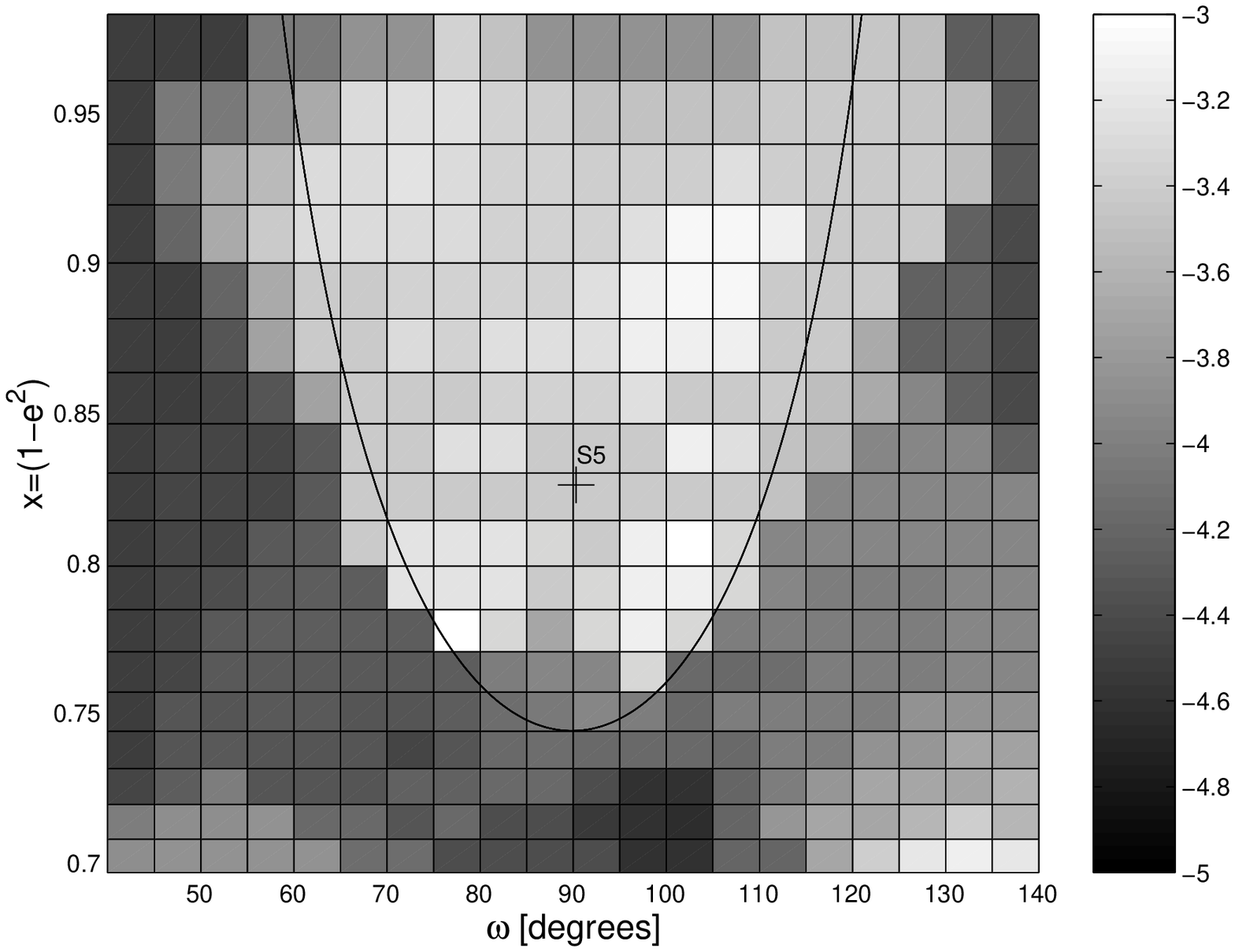}
\caption{Shaded plots of $\sigma$ when the Great-Inequality's period 
was (left) 640 yrs, and (right) 480 yrs.  The position of the secondary 
resonance shifts upward toward the center of the libration island 
in the second case (480 yrs is the minimum period of libration).}
\label{fig: Great_Inequality_change}
\end{figure}

The average and minimum values of $\omega$'s libration periods for 
particles in Kiviuq's region are 640 yrs and 480 yrs, respectively.  
Presently, the Great-Inequality resonance has one location,  
just outside the separatrix, in the region of circulating particles.
An interesting question would be to see what happened when the 
Great-Inequality resonance had a shorter period.  In particular, when 
the value of the Great Inequality was shorter than $\simeq820$ yrs (the 
maximum period of libration), a new secondary resonance should appear 
in the region of librating orbits.  When the Great-Inequality's period was 640 yrs, several particles 
in the libration island would have been in the region of this new secondary 
resonance.  This can be seen in Fig.~\ref{fig: Great_Inequality_change} 
which shows the results of 
the integration of our low-resolution survey integrated under the influence 
of Saturn and a Jupiter whose orbit was modified so that the period 
of the Great Inequality was a) 640 yrs, and b) 480 yrs.
We called these simulations ``static integrations'', as opposed to 
other simulations to be discussed in the next section for which the 
period of the Great Inequality is not fixed but changes with time.
Note how the position of the high-chaoticity region, that 
our simulations show to be connected to the 1:1 resonance 
between $\omega$ and the Great Inequality, is displaced upward 
with respect to the traditional integration for the first case, and is 
very close to the center of the libration island for the second 
case.  More important, a significant fraction (25\%) of orbits that stayed 
inside the libration island in the integration with the present 
configuration of the planets were on 
switching orbits when the period of the Great Inequality was 640 yrs.

\begin{figure}
\plotone{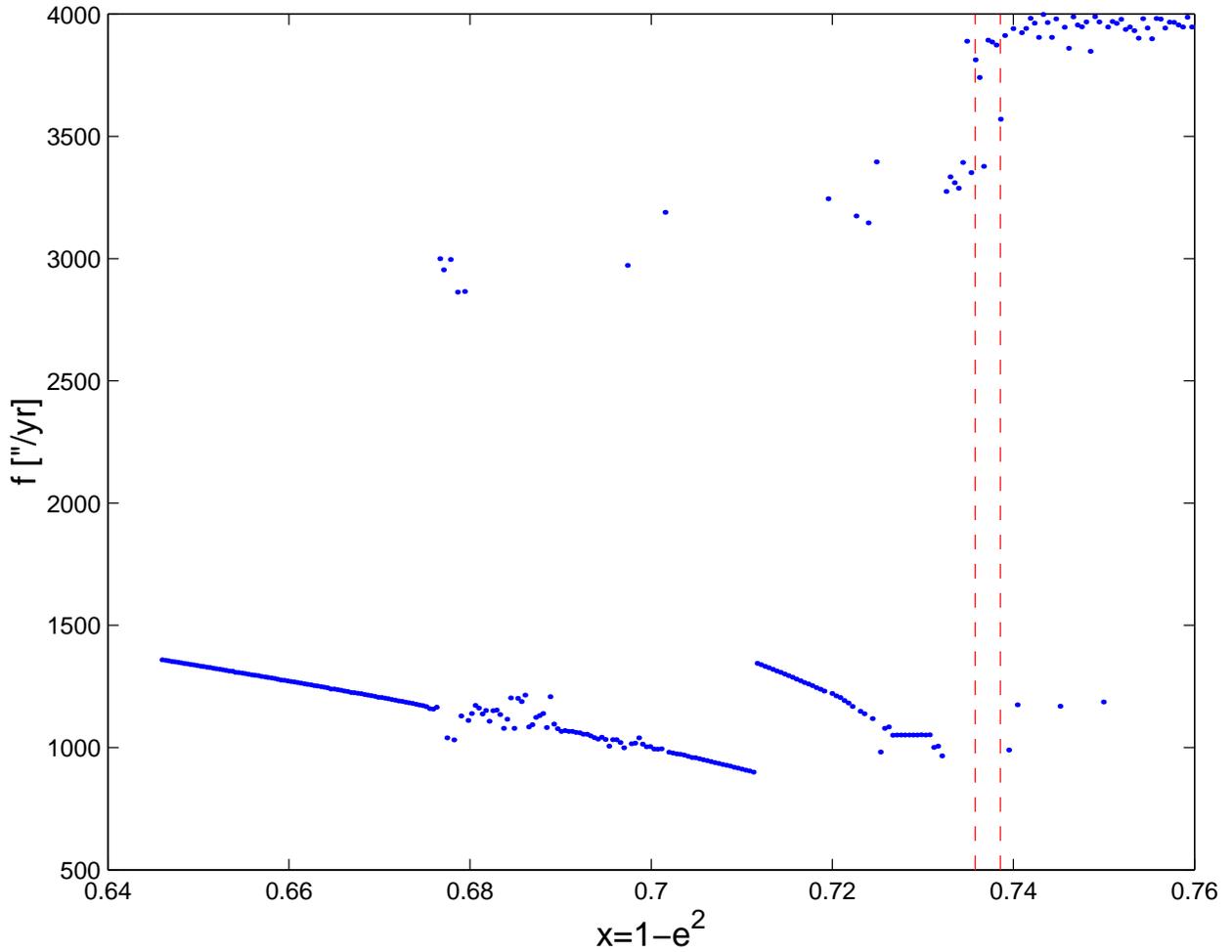}
\caption{Variations in the frequency with largest amplitude as a function
of $x$.  When a resonance is encountered, frequency values are scattered
($x = 0.68-0.69$).  The curve is not continuous at $x = 0.71$ 
because different frequencies have their largest values
of amplitude for different regions in $x$.  The dashed vertical lines show 
the location of the transition region, where orbits switch behavior from
circulation to libration.
\label{fig: frequency_x}}
\end{figure}

These simulations show that indeed a 1:1 resonance between 
$\omega$ and the GI (and the other terms needed 
to satisfy the d'Alembert rules) was present in the libration region 
when the GI's period was lower.
Coming back to our problem of identifying secondary resonances for the 
high-resolution survey, the method applied to our low-resolution survey 
to change the GI's period can give precious insights on the 
identity of those resonances.   Fig.~\ref{fig: frequency_x} shows
how the frequency with largest 
amplitude in the spectra obtained with FAM (just the frequency with
largest amplitude, not 
necessarily the frequency associated with $\omega$'s precession) for 
our high-resolution survey 
changes with $x$. When a secondary resonance is encountered, values of the 
frequencies spread around rather than following a quasi-linear behavior
(the feature associated with the pericentric secular resonance at $x=0.68$ 
is most instructive).

\begin{figure}

\epsscale{.45}
\plotone{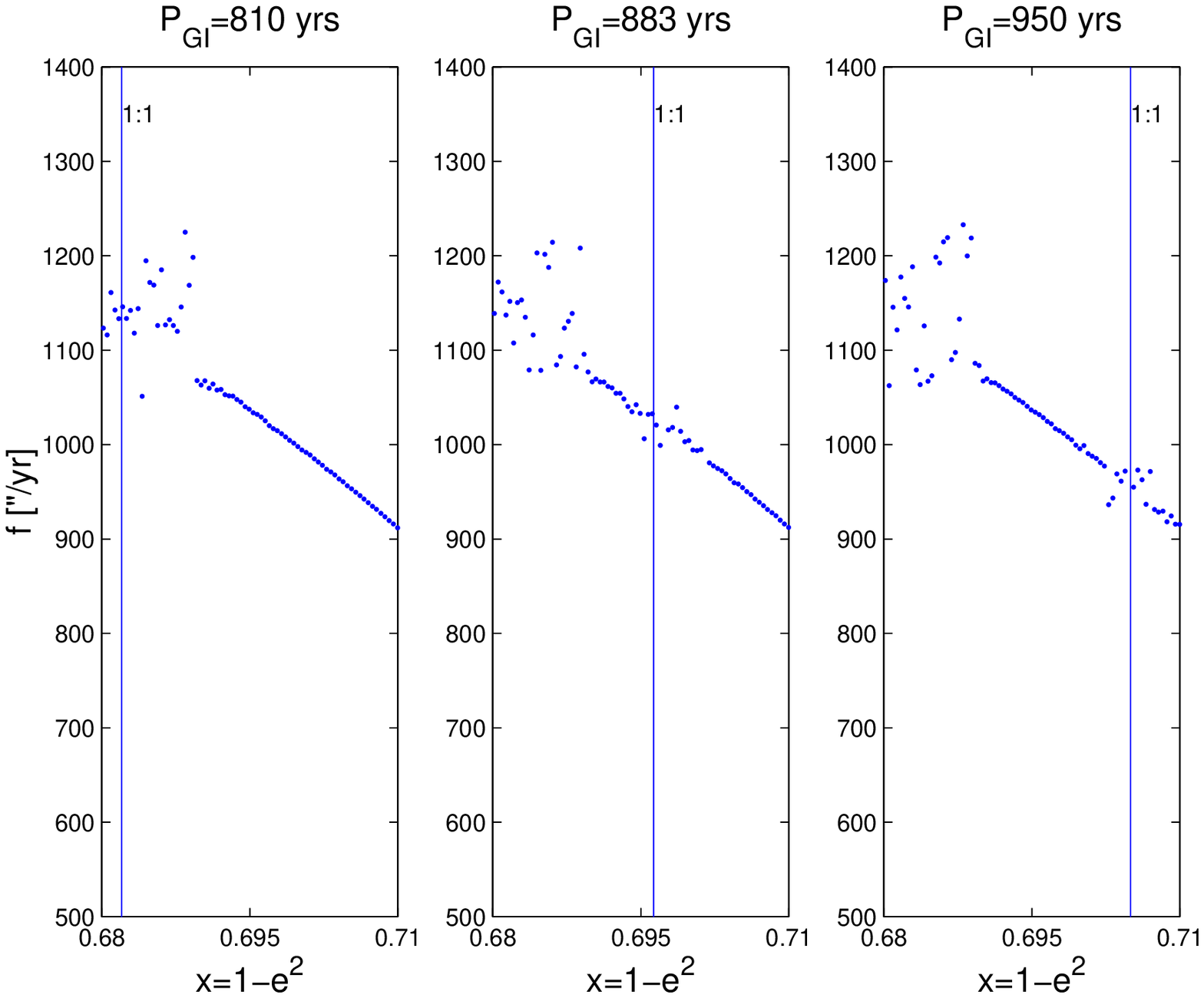}

\epsscale{.45}
\plotone{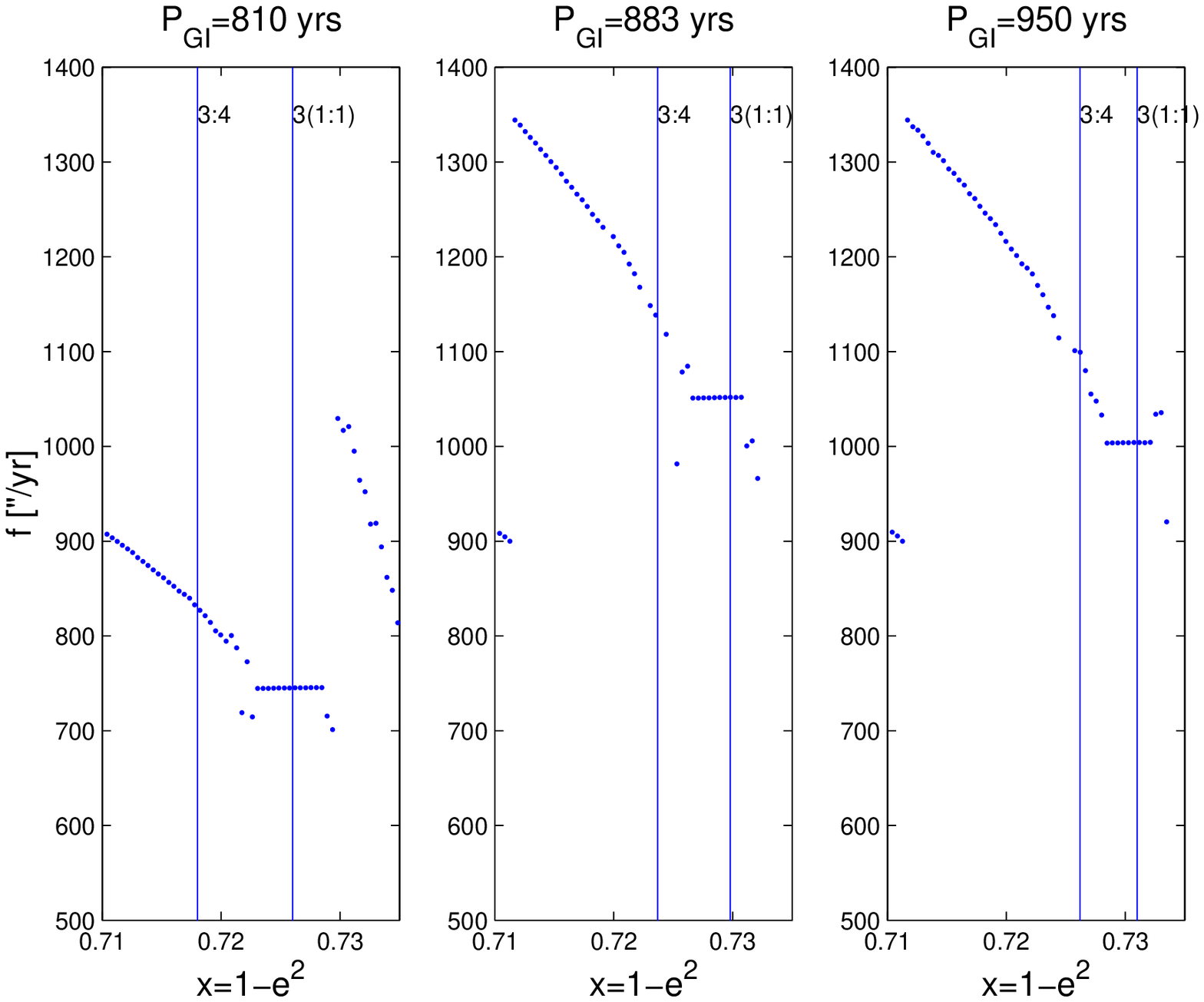}

\epsscale{.45}
\plotone{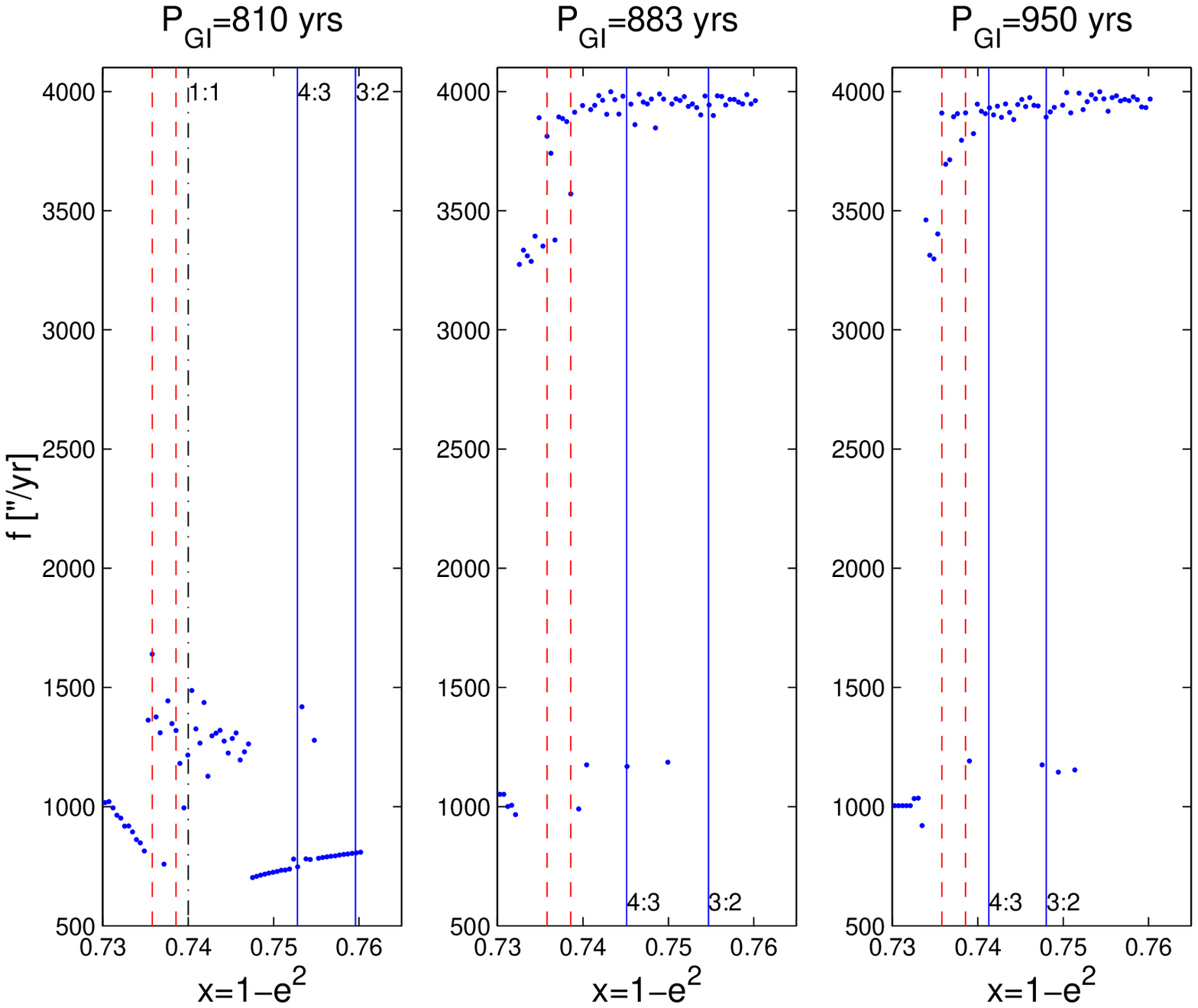}

\caption{Variations in the frequency having the largest amplitude as a function
of $x$, for three regions in $x$, and three values 
of the Great-Inequality's period (810, 883, and 950 yrs, respectively).  
The first row of figures reports the region around 
the 1:1 GI resonance, the second the region around the 3:4
and resonance, and the third 
the region of librating particles.  Vertical lines show the expected position
of the resonance (identified at the top of the line) based on its 
resonant argument.  By 3(1:1) we indicate the resonance of argument 
$3(\varpi-\Omega-5\lambda_S+2\lambda_J+3\Omega_S)+(\Omega-\Omega_S)$. 
Dashed vertical lines in the third row of figures
show the position of the transition region.  See text 
for more details. 
\label{fig: frequency_x_3}}
\end{figure}

To test our interpretation of the secondary resonances, we
modified Jupiter's position so that the GI period became  
810, 883 (current value), and 950 yrs, 
respectively, and integrated the high-resolution survey 
with the three configurations.  Figs.~\ref{fig: frequency_x_3}a, b, and c 
illustrate what happens to the frequencies
for these three values of the GI, for the region 
of the 1:1 resonance, the 3:4, and for librating particles, respectively. 
Fig.~\ref{fig: frequency_x_3}a shows the behavior in the region of the 1:1 resonance.  The
resonance position (computed using Eq:~\ref{eq: f_resonant} and the 
different GI periods)
shifts toward larger $x$ when the GI period
becomes longer.  When the period equals 810 years,
the position of the resonance corresponds to that of the pericentric
secular resonance (note how its position does not move when the 
GI period is changed), and it goes to higher values of $x$
when we increase the period.  The vertical lines indicate the expected 
position of the resonance when the GI period is modified.
The fact that 
there is an excellent agreement between the predicted position of the resonance
and the results of our numerical simulation seems to further confirm 
our hypothesis for the source of chaos in this region.

The middle row of 
Fig.~\ref{fig: frequency_x_3} 
shows the same plots, but for $x=0.71-0.735$.  This 
region contains two resonances of argument 
$4(\varpi-\Omega)-3(5\lambda_S-2\lambda_J)+8\Omega_S+\varpi_S$, and
$3(\varpi-\Omega-5\lambda_S+2\lambda_J+3\Omega_S)+(\Omega-\Omega_S)$: 
these are given the short-hand notation of 4:3 and 3(1:1).
Once again, the numerical simulations demonstrate that the chaos zones
move with the resonances, i.e., they agree with the 
predictions based on our resonant arguments (vertical lines).

Finally, the bottom row of Fig.~\ref{fig: frequency_x_3} 
shows the values of the frequencies of largest amplitude for the region 
of librating satellites.  The dashed vertical lines represents the transition
region in which particles alternate from circulating to librating behavior.
The left panel shows substantial frequency diffusion
near the separatrix, 
when $P_{GI}=810$ yrs, not observed for longer Great-Inequality periods.
We believe that this is due to the appearance of the 1:1 resonance 
(having argument $\varpi-\Omega-5\lambda_{S}+2\lambda_{J}+3\Omega_{S}$)
in the region of librating particles (see the vertical line on the plot).
This resonance is simply out of the range of periods in $\omega$ for 
librating particles when $P_{GI}> 820$ yrs. 

Regarding other resonances in the region, we are still limited by the 
problem of plotting the resonant argument.  However, our simulations suggest
at least two other resonances in this region:
$2(\varpi-\Omega)-3(5\lambda_{S}-2\lambda_{J})+8\Omega_{S}+\varpi_S$ and 
$3(\varpi-\Omega)+4(5\lambda_{S}-2\lambda_{J})+11\Omega_{S}+\varpi_S$.
The bottom row of 
Fig.~\ref{fig: frequency_x_3} shows how the expected positions of these 
resonances shift when the Great-Inequality period is modified.  In each 
case, the expected position is very close to regions of frequency variation.
We believe this is strong circumstantial evidence in favor of the existence of 
these resonances.  Other resonances containing combinations
of only $\varpi_J$ or $\varpi_S$ instead of $\Omega_S$ (plus any other 
term needed to satisfy the second d'Alembert rule) in their argument 
are also possible, but,
as observed in other cases, are associated with features of weaker chaos 
(their locations are not shown in Fig.~\ref{fig: frequency_x_3}). 
Regarding the bottom row of Fig.~\ref{fig: frequency_x_3}, we observe 
that the resonance positions shift toward the separatrix as 
the GI period increases.

To conclude, we have identified several secondary resonances connected to the 
Great Inequality for the region near the separatrix of the Kozai resonance.
The likelihood that the jovian planets had different past positions 
means that the locations of these resonances might have also been different.
This introduces interesting new prospectives for the stability of primordial
satellites inside the Kozai resonance, that we now investigate.

\section{Effects of Planetary Migration on Any Primordial Population
of Satellites in the Kozai Resonance}

We have just argued that a web of secondary resonances 
involving $\omega$, the argument of the Great Inequality, and other terms
exists in the region of phase space around the separatrix between libration 
and circulation.
Considering that the period of the Great Inequality might have been different 
in the past, we ask ourselves if  
a mechanism of sweeping secondary resonances inside 
the Kozai resonance, not dissimilar from the one that acted in the Kuiper 
Belt, could also have acted inside the Kozai resonance.  If that is 
the case, what are the repercussions for the stability of satellites' orbits
in the Kozai resonance?

\begin{figure}
\epsscale{1.10}
\plottwo{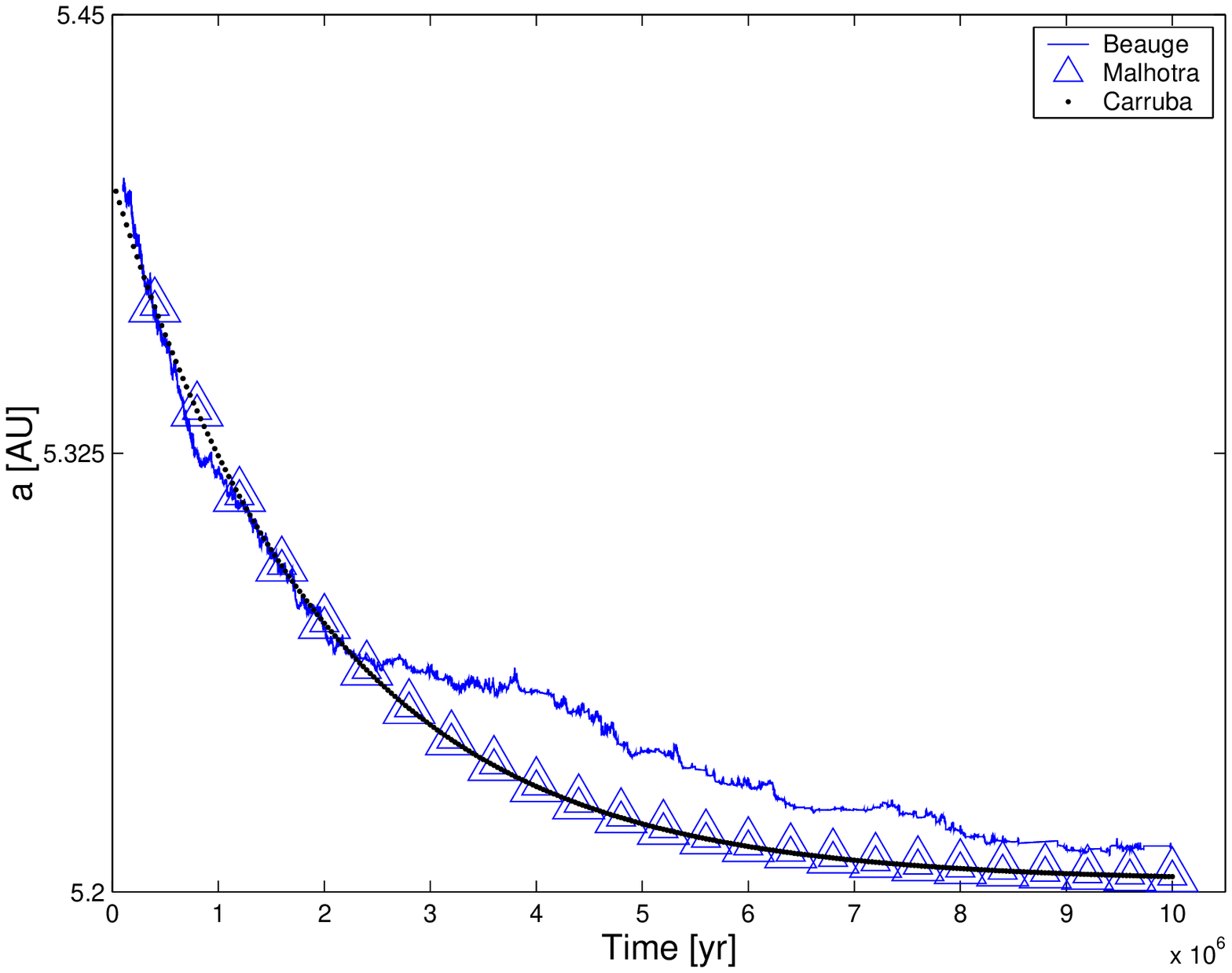}{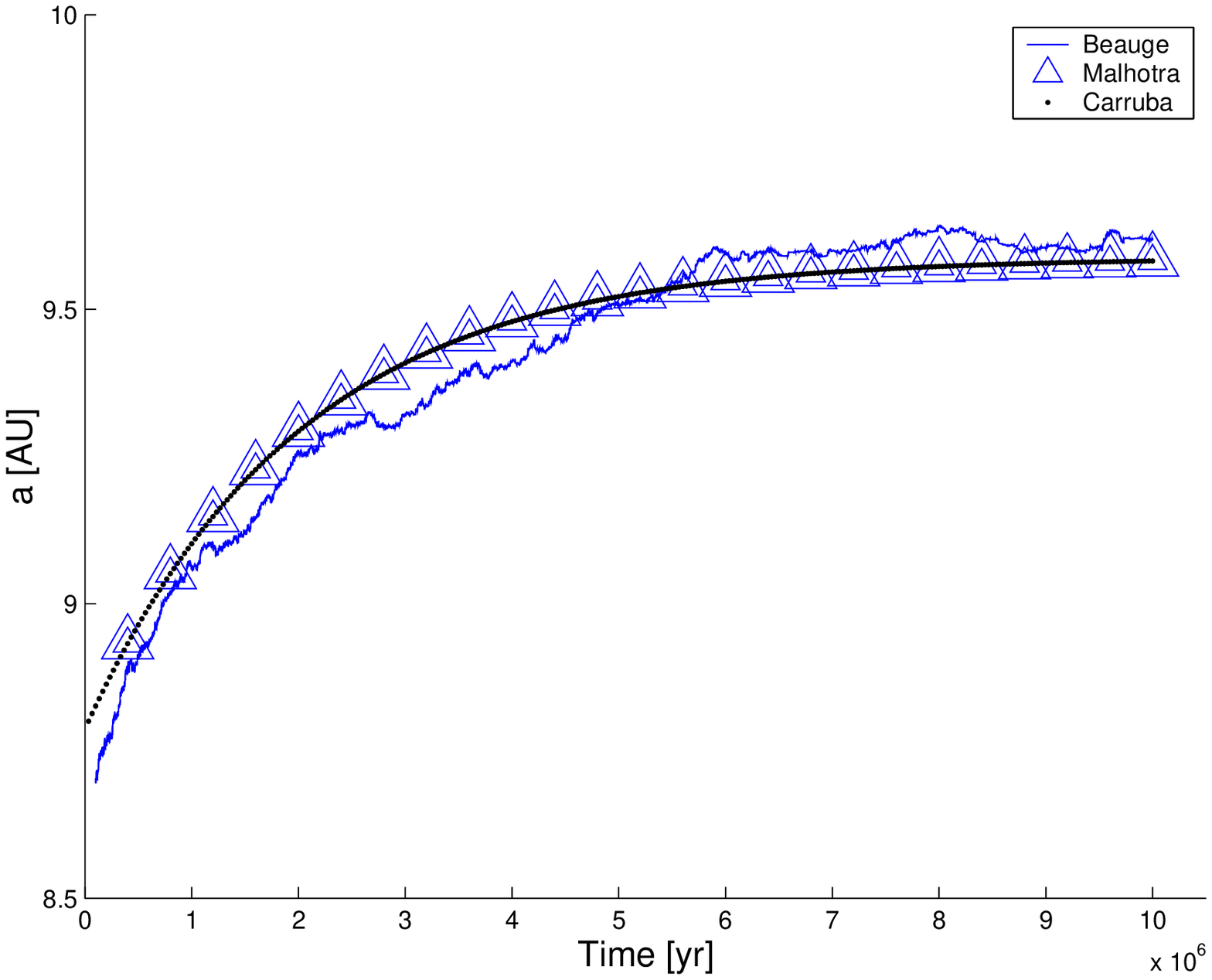}
\caption{The time evolution of the
semimajor axes of (left) Jupiter and (right) Saturn.  The black band plots 
the results of our integrator, the triangles show the result of the 
integration of Eq.~\ref{eq: a_vs_t}, and the line displays the results of 
a simulation on planetary migration from Beaug\'{e} {\em et al.} 2002, 
as described in the test.}
\label{fig: Planet_migr}
\end{figure}

To address these questions we need to a) have a model of planetary migration
and b) have an integrator able to simulate the effect of planetary migration 
for Jupiter and Saturn.  Following Malhotra (1995), we assumed 
that the semimajor axis varied as:

\begin{equation}
a(t)=a_{f}-\Delta a \exp{(-t/\tau)},
\label{eq: a_vs_t}
\end{equation}

\noindent where $a_f$ is the semimajor axis at the current 
epoch, $\Delta a$ is 
the change in semimajor axis (equal to -0.2 AU for Jupiter and to 
0.8 AU for Saturn) and $\tau$ (= $2 \cdot 10^6$) is 
a characteristic timescale for 
migration.
For the integrator to simulate planet migration, 
we followed this recipe.  We modified SWIFT-WHM so that an additional 
drag-force was applied to each planet along the direction of  
orbital velocity (by the symbol {\bf $\overline{v}$} we 
identify an unit vector along the direction of orbital velocity).  
This produces an acceleration:

\begin{equation}
\Delta {\bf \dot{v}} = \frac{\bf{\overline{v}}}{\tau}\left(\sqrt{\frac{G_0 M_{sun}}{a_f}}-
\sqrt{\frac{G_0 M_{sun}}{a_i}}\right)\exp{(-t/\tau)},
\label{eq: planet_drag}
\end{equation}

\noindent where {\bf \.{v}} is the acceleration and $a_i$ is the initial 
position of a planet (Malhotra 1995).  We neglected the effect of the
planetesimals' perturbations on the satellites.  
Fig.~\ref{fig: Planet_migr} displays several computed evolutions of 
Saturn's and Jupiter's semimajor axes.
The two curves agree well.  These are to be compared to the results of 
Beaug\'{e} {\em et al.} (2002) who 
used SWIFT-RMVS3 to simulate the evolution of a disk 
of 1000 massless planetesimals subjected to the gravitational perturbations 
of the four major outer planets.  The evolution of the planet's 
semimajor axes generally 
follows the exponential law of Eq.~\ref{eq: a_vs_t}, but in addition is 
affected by short-time variations
due to the impulses of single encounters.

Having now the tools for our investigation, we must define 
initial conditions for our simulations.  To investigate the effect of 
planetary migration on the stability of orbits in the Kozai resonance, 
we first integrated Kiviuq with the integrator
that accounts for planetary migration backward in time for 10 Myr.
Not surprisingly, Kiviuq remained inside the Kozai resonance 
for the full length of the integration. We then used the last 
100,000 yrs of the integration to compute the averaged orbital elements 
of Kiviuq and $\Theta$'s value (= 0.4322).  Using this information, 
we generated a low-resolution survey of test particles, exactly as in Sec. 2.  
We then integrated this new set of initial 
conditions forward in time for 10 Myr.  During this forward integration, the 
Great-Inequality's period increased from 90 yrs at the beginning
of the simulation to the current value of 883 yrs at the end of the 
integration.

\begin{figure}
\epsscale{1.10}
\plottwo{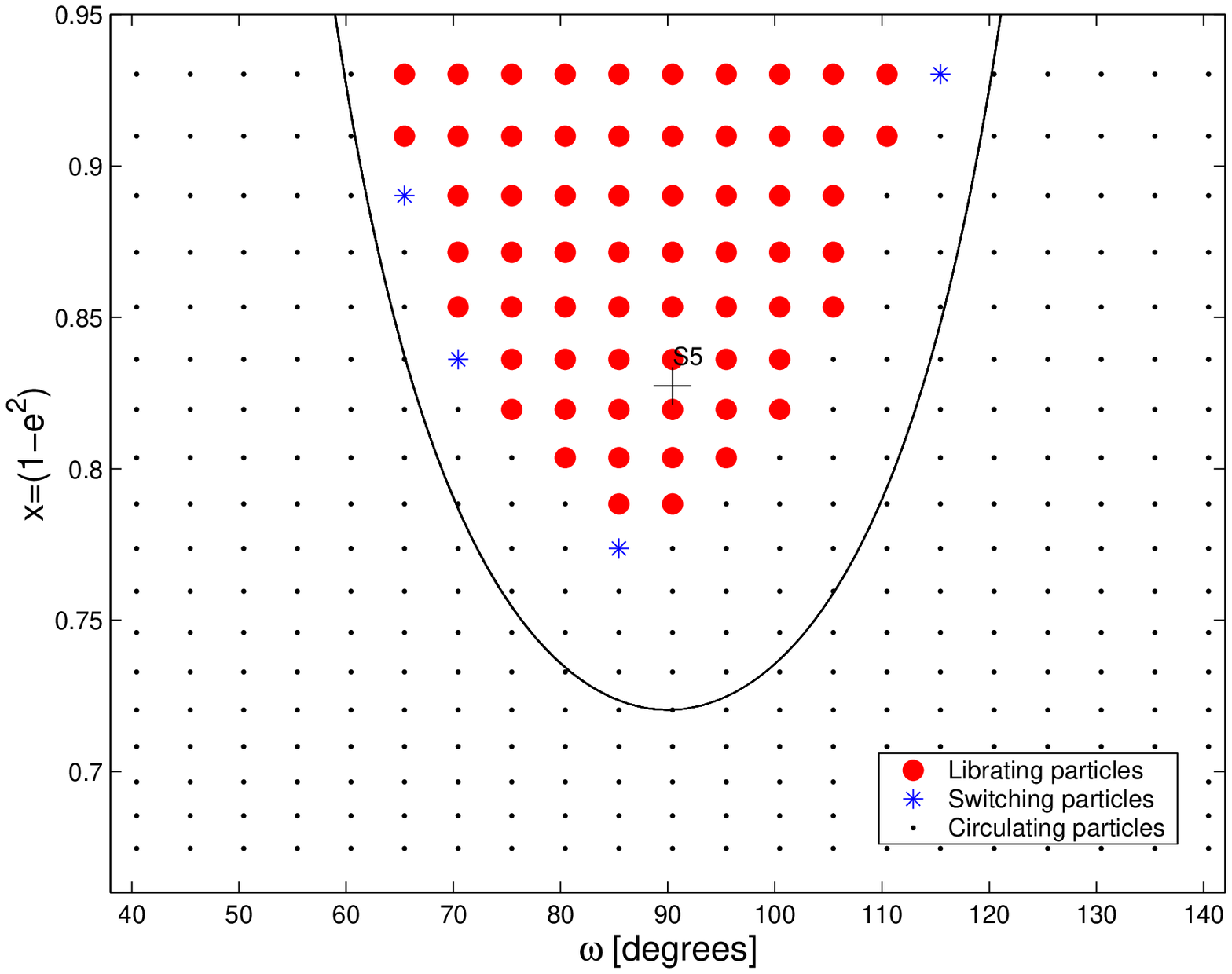}{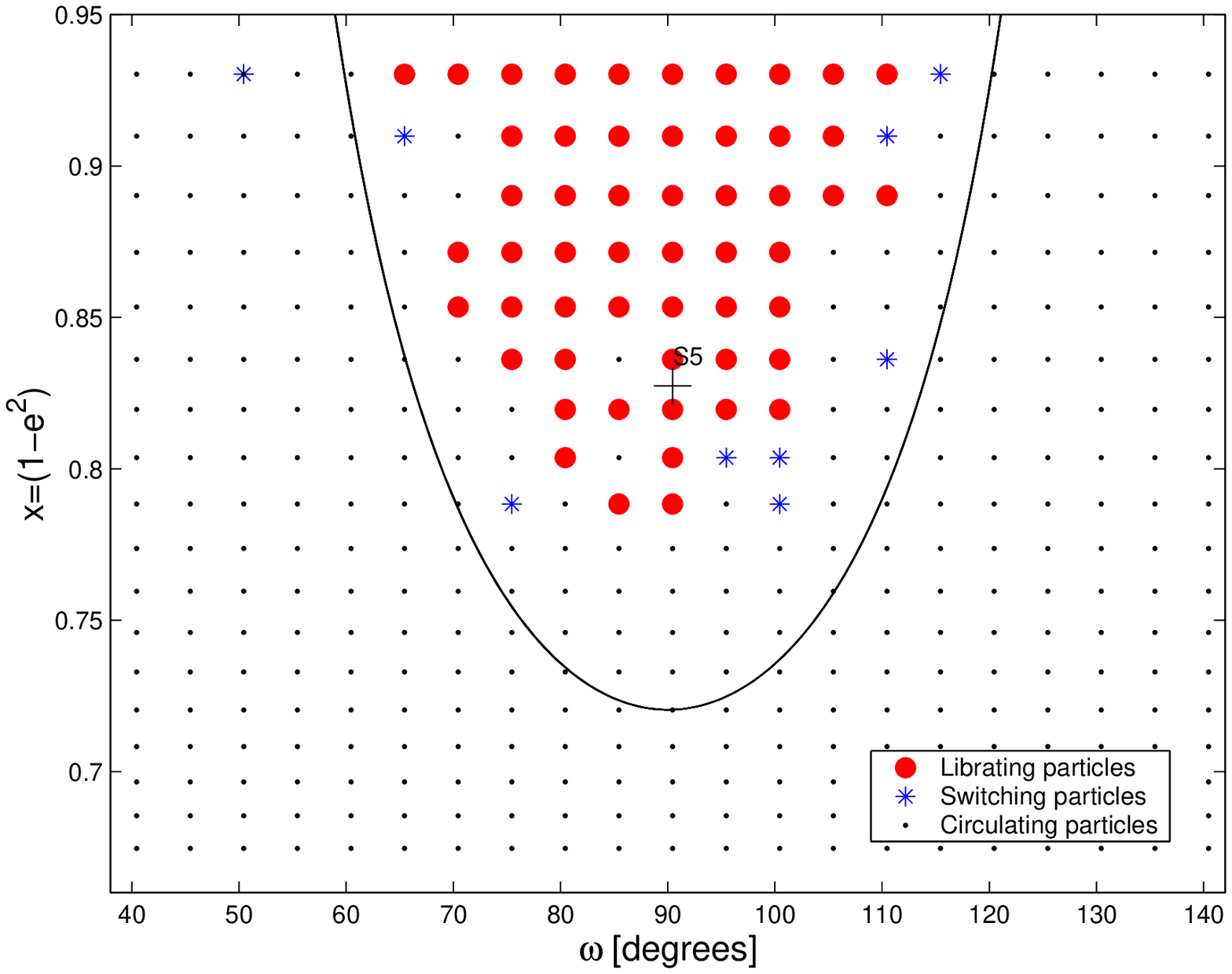}
\caption{Fate of test particles for our low-resolution survey, where 
we used Eq.~(\ref{eq: a_vs_t}) to simulate planet migration.
a) gives the orbital nature at the simulation's beginning, 
while b) shows it at $t$ = 5 Myr; the symbols are the same used in 
Fig.~\ref{fig: particles_fate}.  At the end of the 
simulation only one additional particle was lost from the libration island.}
\label{fig: Planet_migr_particles_fate}
\end{figure}

Fig.~\ref{fig: Planet_migr_particles_fate}
shows the particles' fates at left $t$ = 0 and right $t$ = 5 Myr.  14\% of the 
particles originally in the libration island were on circulating or switching 
orbits by $t$ = 5 Myr.
At the end of the simulation, 15\% of the test particles originally 
in the libration island were no longer Kozai resonators.  The 
mechanism of sweeping secondary resonances seems to be effective  
in depopulating the resonance.  This mechanism not only 
affects orbits near the separatrix, but also those particles well inside 
the Kozai resonance.  Examples are the particles with ${\omega}_0$ = 
$85^{\circ}$ and $x_0$ = 0.80 and 0.83, which are 
very close to Kiviuq's orbital region.  This shows that this region 
is also affected by the sweeping secondary resonances. 

\begin{figure}
\epsscale{0.99}
\plotone{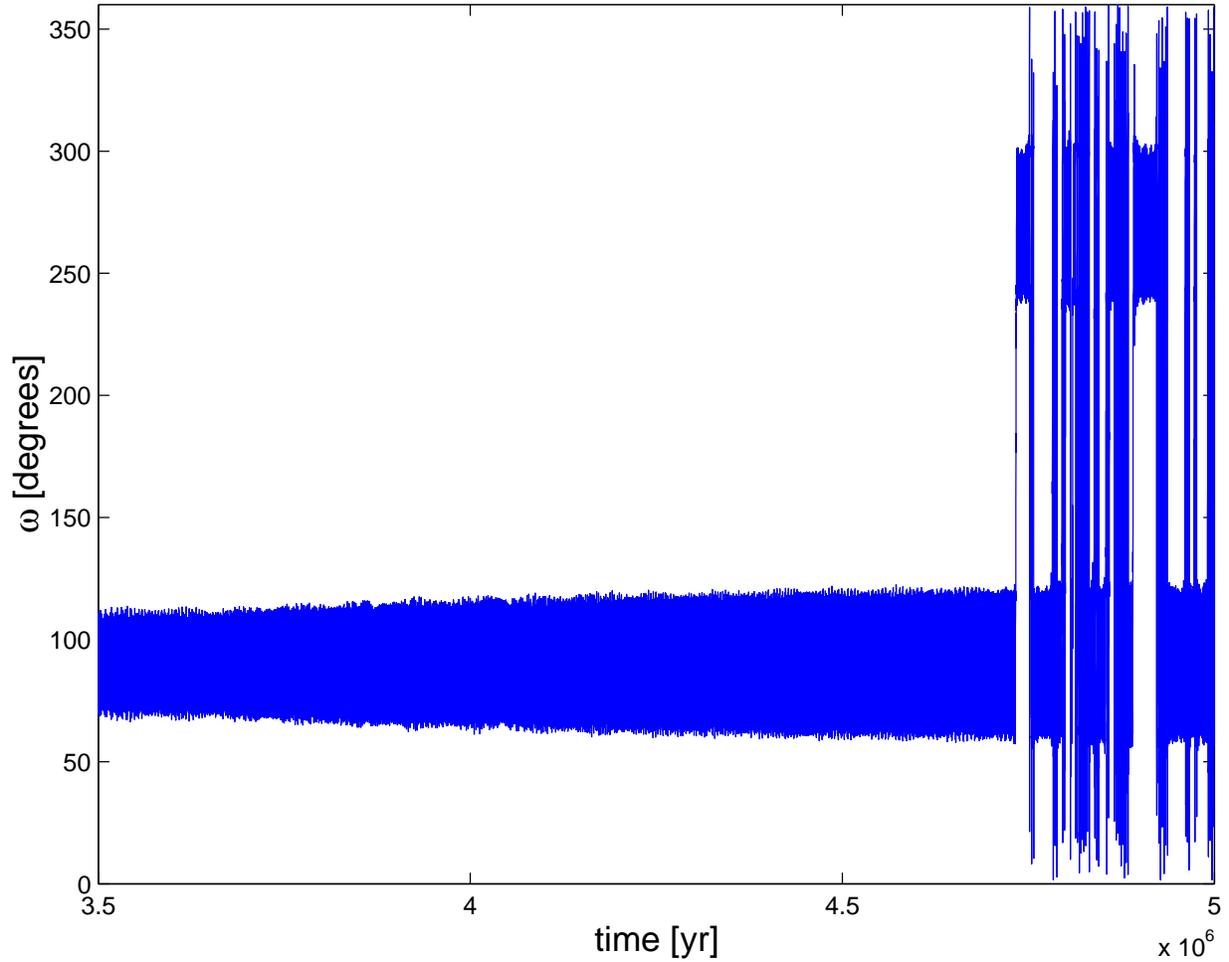}
\caption{Time evolution of $\omega$ for an orbit initially inside the 
Kozai resonance ($x_0=83,{\omega}_0=85^{\circ}$).  At $t =3.7$ Myr the 
particle is captured into the secondary resonance with the Great Inequality.
As a consequence, the amplitude of libration slowly increases until the 
test particle reaches the separatrix of the Kozai resonance. At that point, it
escapes from the resonance and starts switching 
back and forth from circulation to libration (around either $90^{\circ}$ or 
$270^{\circ}$).  At the end of the integration
($t = 10$ Myr) the particle was on a circulating orbit.}
\label{fig: omega_period_50}
\end{figure}

We explain these results in 
the following way:  when, owing to the planets' 
migration, the secondary resonance's period approaches 
$\simeq400$ yrs, a few of the particle inside the Kozai resonance (but 
only a fraction, the process is not 100\% efficient) are captured into 
a secondary resonance, 
and move further out in the libration island until the secondary-resonance's 
period attains $\simeq700$ yrs.  For such libration periods, the orbits are  
so close to the separatrix that escape becomes possible.  This mechanism seems 
to be confirmed by the time evolution (Fig.~\ref{fig: omega_period_50})
in $\omega$ for one of these particles ($x_0 = 0.83,{\omega}_0 = 85^{\circ}$).

The fact that most particles
escape from the Kozai resonance 
in the first half of the simulation is not surprising.  
The period of the Great Inequality reaches 600 yrs during the 
first 5 Myr, i.e., the average value of the precession period for 
orbits in the 
Kozai resonance (see Fig.~\ref{fig: Periods}).  After 5 Myr, the growing 
value of the Great-Inequality period sweeps fewer and fewer 
particles inside the Kozai 
resonance, and the mechanism loses its efficacy.

Another interesting consideration is that not only librating particles
are pushed toward the separatrix, also some circulators in general are 
captured by one of the secondary resonances and are carried toward the 
separatrix.  The fact that circulators do not generally become librators
is due to the fact that in order to cross the separatrix they have to reach 
very high values of the precession period in $\omega$ 
(see Fig.~\ref{fig: omega Periods}).  Once close to the separatrix, they reach 
a highly chaotic region, and many particles are lost from the secondary 
resonances.  This mechanism could explain why so many irregular 
satellites are currently found in the proximity of the Kozai separatrix
(\'{C}uk \& Burns 2004 for \aj).

A word of caution should be given about our results. We 
only used a smooth 
exponential evolution (Malhotra 1995) for the planet's  
migration. In the real Solar System things were most likely
different from this model; e.g., Hahn and Malhotra (2000)
introduced short-time scale variations in Neptune's outward expansion
by adding some random jitter to the torque applied to Neptune (cf.
the $a$ history by Beaug\'{e} {\em et al} (2002) shown in 
Fig.~\ref{fig: Planet_migr}).
This jitter was parametrized by the standard deviation 
${\sigma}_{jitter}$ of the planet-migration torque in units of the 
time-averaged torque.  For small values of ${\sigma}_{jitter}$ ($< 10$), 
capture efficiencies in the 2:1 resonances were substantially unaltered, 
and only when ${\sigma}_{jitter}$ reached values larger than 25 were capture 
probabilities reduced. 

In our case, we can assume that a similar mechanism should be at work, 
and that any significant amount of short-time variations in the semimajor 
axes of both Jupiter and Saturn might in principle reduce the capture 
efficiency into the Great-Inequality secondary resonance.
Since our work is, in many ways, exploratory, we believe the use of 
the exponential model is defensible at this stage of our study.  But we 
acknowledge that a more realistic model for the motions 
of Jupiter's and Saturn's semimajor axes should be used, 
before realistic values of capture probabilities for our mechanism of
sweeping secondary resonances could be computed.

In any case, we believe that our mechanism of sweeping secondary 
resonances might be an effective scheme
for destabilizing the primordial population of bodies originally in the Kozai
resonance.  The mechanism should not be 
limited to Kiviuq's region.  The fact that the Great-Inequality's 
period swept through values starting from 90 yrs at the beginning of the 
simulation with planetary migration to the current 883 yrs, 
implies that an analogous 
mechanism should also have been at work in the regions of the other Saturnian 
and Jovian satellites presently in the Kozai resonance, for $\omega$ libration 
periods of less than 1000 yrs.
 
\section{Summary and Conclusions}

We have analyzed the orbital behavior of satellites inside and outside the 
Kozai resonance for the phase space around the Saturnian satellite 
Kiviuq.  Our goal was to identify areas of chaotic behavior and to understand
if chaotic diffusion might have played an important role, now or in the past, 
for orbital stability inside the Kozai resonance.
By applying the frequency analysis method (FAM) and computing the maximum 
Lyapunov exponents, we identified several secondary 
resonances.  The major source of chaos for orbits in the 
circulating region are secondary resonances
involving the argument of pericenter $\omega$ (= $\varpi-\Omega$), 
the Great-Inequality's argument
($2{\lambda}_{J}-5{\lambda}_{S}$), the longitude of the node 
($\Omega$),
and the planetary secular frequencies ($g_5, g_6$, and $s_6$).  The region 
of transition between circulation and libration of $\omega$ 
is strongly chaotic, when 
perturbations from the other jovian planets are considered.
According to our secular model, variations in Saturn's semiminor axis 
(and, therefore in its eccentricity and semimajor axis) are what drive
the chaotic behavior, especially the transfer from circulation to libration.
While secular variations modulated by the $g_5, g_6, $ and $s_6$ frequencies
are not negligible, once again our results indicate that the dominant effect
is connected with perturbations having the Great-Inequality's period.

Also, contrary to what might be 
expected from the secular model, our results indicate 
that the chaotic layer at the boundary between circulation and libration 
is not symmetric, but shows an asymmetry between values of $\omega$ larger 
and smaller than $90^{\circ}$.  This is a consequence of other perturbations
connected with secular pericentric resonance, evection 
inequality, and variations in 
Saturn's semimajor axis, which are strongly dependent on the initial 
conditions for particles' $\varpi$ and 
the Sun's position.  Our simulations suggest that the dominant 
effect is connected with the secular resonance, but variations in Saturn's 
semimajor axis are not entirely negligible.

From our simulations we can introduce the following 
scenario for chaotic orbits near Kiviuq's Kozai resonance.  Starting 
from librating 
orbits close to the libration center, we possibly identify two weak 
secondary resonances of argument 
$2(\varpi-\Omega)-3(5\lambda_S-2\lambda_J)+8\Omega_S+\varpi_S$ and 
$3(\varpi-\Omega)-4(5\lambda_S-2\lambda_J)+11\Omega_S+\varpi_S$
that introduce chaotic behavior.  The first strong 
source of chaos is however connected with the transition region between 
circulation and libration: orbits start to switch back and forth from 
circulation to libration and the whole region is characterized by high 
values of $\sigma$.  Once in the circulating regime, a strong resonance 
of argument 
$3(\varpi-\Omega-5\lambda_S+2\lambda_J+3\Omega_S)+(\Omega-\Omega_S)$
appears at $x$ = 0.73.
In addition to that, the dominant source of chaos, excluding the nearby 
pericentric secular resonance, is connected to a secondary 
resonance with the Great Inequality of argument 
$\varpi-\Omega-5\lambda_{S}+2\lambda_{J}+3\Omega_{S}$ 
(these orbits have argument-of-pericenter 
periods of order 840 yrs).  Orbits in the region of this secondary 
resonance show some of the highest $\sigma$ values among the test particles we 
studied. Other secondary resonances connected with the Great Inequality 
are also observed.

The fact that the dominant source of chaos for orbits in Kiviuq's region is connected with the Great-Inequality secondary resonance opens 
interesting perspectives.  It is believed that jovian planets
formed in different locations than their current ones, and then migrated by 
gravitationally scattering the primordial population of planetesimals
(Malhotra 1995).  The possibility that planets occupied different 
semimajor axes in the past has important repercussions, since the Great-Inequality's period is connected with the osculating semimajor
axes of Jupiter and Saturn.  In particular, when the two 
planets were closer together,
the period of the Great Inequality was shorter.  We simulated this 
effect first by changing the position of Jupiter 
so that the Great-Inequality's period was 640 yrs and 480 yrs,
respectively (``static integrations'').  These simulations showed that, 
indeed the position of the Great-Inequality's 
secondary resonance shifted higher.  A twin 
resonance for librating particles appeared when the Great-Inequality's period equaled the average libration period (640 yrs), 
and the minimum (480 yrs) (this resonance is not visible today 
because the maximum period of libration inside the Kozai resonance 
is of $\simeq 820$ yrs).  Then we fully integrated Jupiter and Saturn 
with an integrator that simulates planet migration according to 
the exponential law of Malhotra (1995).

Our results show that a mechanism of sweeping resonances,
analogous to the one that operated in the Kuiper Belt (Malhotra 1995), 
or inside the 2:1 mean motion resonance with Jupiter (Ferraz-Mello 
{\em et al.} 1998) must have operated as well inside the Kozai resonance.
Our simulations show that 15\% of the satellites originally inside 
the Kozai resonance became circulators or exhibited switching behavior 
when planets acquired their final orbits.
Indeed several particles were captured into the secondary resonance 
with the Great Inequality (or other secondary resonances) 
and, as a consequence, their amplitudes of 
oscillation in $\omega$ increased until the test particles reached
the boundary between circulation and libration, where escape became  
possible.  Several particles on circulating orbits were also pushed 
toward the separatrix by an analogous mechanism of sweeping 
secondary resonances (but none crossed the boundary to become a librator).
We believe that a similar mechanism might have acted in the past 
also for the region of phase space associated with the other jovian satellites 
currently inside or near the Kozai resonance.
This process could explain why so many irregular satellites are currently 
found near the Kozai separatrix (\'{C}uk and Burns 2004).

In this work we used several numerical tools.  In a few cases, like 
for example the median filter for our FAM results, we introduced a technique, 
that, while employed in other fields (for example in remote-sensing and in the 
analysis of data from martian probes), was not, to our knowledge, used 
previously in celestial mechanics for reducing FAM results.
Still on the technical side, we created new integrators to 
account for the Bretagnon solutions of the giant planets and to simulate 
planetary migration (Malhotra 1995).  We believe these tools might be 
useful for other studies,
not only for the irregular satellites of jovian planets.

Several questions remain that might be the subjects of future research. 
In this work we could not compute reliable
capture probabilities for the Great-Inequality 
resonances when planetary migration is considered, since 
short-period variations in the semimajor axes of Jupiter and Saturn 
should also be taken 
into account (Hahn and Malhotra 2000, Beaug\'{e} {\em et al.} 2002).
It would also be interesting to extend our studies to the neighborhoods of the
other jovian satellites currently inside the Kozai resonance, and see if 
a web of secondary resonances (possibly involving the Lesser 
Inequality, the 2:1 quasi-resonance between Neptune and Uranus) 
is also present for the case of those neptunian satellites in Kozai
resonance (Holman {\em et al.} 2004).  We hope that this work might have
opened interesting new prospects for the study of origins and stability
of irregular satellites in Kozai resonance.

\section{Acknowledgments}

We are grateful to Thomas Loredo, Sylvio Ferraz-Mello, Richard Rand, Doug 
Hamilton, and Suniti Karunatillake for useful comments and suggestions supported in part by NASA's Planetary Geology and Geophysics program.
More information about this work is available at:

\url{http://www.astro.cornell.edu/~valerio/FFT}.

\appendix

\section{Median filter and ${\chi}^2$ test}

When measuring frequencies with a FMFT, the shape of 
the frequency peaks can be modeled as Gaussian, but 
our measurements are made at isolated points.  This introduces what in signal 
processing is called ``impulse noise'', which can be attenuated by the use 
of a median filter (Press {\em et al.} 1996, Pitas 2000). 

In our case we took data on a running 3x3 matrix; the value at the central
point is the median of the nine elements.  Then we moved the matrix by one 
element and repeated the procedure.  To compute values of $\sigma$ for 
points along the boundary of our figure, we added two external rows 
and columns of points all with values of $\sigma$ obtained from the 
average of all points in the figure.
Once all elements were computed, we 
calculated the average of the percentage difference between the values of 
$\sigma$ before and after the median filter was applied.   If the 
percentage difference was larger than 5$\%$, we repeated the procedure until
convergence was assured.   To test that our median filter is actually removing 
only noise and not actual data, we devised the following  test.
We applied our median filter to one of our low-resolution surveys, until 
the 5\% convegence criterion was fulfilled.    We then computed the difference 
${\sigma}_{ij}$ between filtered ($F_{ij}$) and raw data ($R_{ij}$), and 
computed the standard deviation of the difference $\overline{\sigma}$.
We then generated a fictitious matrix ${{\sigma}_{ij}}^{'}$ of noise following 
a Gaussian distribution with standard deviation $\overline{\sigma}$ and 
zero mean, and added this fictitious noise to the filtered data, obtaining 
a matrix of $\sigma$ values $R_{ij}^{'}$.  The median filter was then reapplied
to these new data until convergence was assured, and a new matrix $F_{ij}^{'}$
obtained.

To compare the two distributions $F_{ij}$ of filtered data and $F_{ij}^{'}$ 
(data obtained after the filtration of the fictitious noise), we devised the 
following scheme.  We computed a ${\chi }^2$ variable, whose expression is 
given by:

\begin{equation}
{\chi}^2={\Sigma}_{i} {\Sigma}_{j} \frac{({F_{ij}}^{'}-F_{ij})^2}
{{\sigma_{ij}}^2+{{{\sigma}_{ij}}^{'}}^2},
\label{eq: Chi^2}
\end{equation}

\noindent where $\sigma_{ij}$ and ${\sigma_{ij}}^{'}$ are the errors on the values of 
the filtered data, intended as the difference between raw and filtered data.
The variable so computed is supposed to follow a ${\chi}^2$ -like distribution.
The problem is to determine the number of degrees of freedom of the 
distribution.   This is not just given by the number (399) 
of data points in our 
low-resolution survey since, when taking the median, each value 
of $\sigma$ is connected to the value of at least some of its 
eight neighbors.  Moreover, the process of filtering is applied 
several times, until convergence is assured, so that the number of 
degrees of freedom can be substantially smaller than the number of 
data points, in a way that is very difficult to estimate.

However, as a rule of thumb, a value of ${\chi }^2$ considerably 
smaller than 399 is a relatively good measure of a good fit.  In the case 
of our test, the value of ${\chi }^2$ was 55, so we believe 
our test should be a convincing proof of the validity of our 
application of the median filter.


\clearpage

\clearpage

\clearpage

\clearpage

\clearpage

\clearpage

\clearpage

\clearpage

\clearpage

\clearpage

\clearpage

\clearpage

\clearpage

\clearpage

\clearpage

\clearpage

\clearpage

\clearpage

\clearpage

\begin{small}
\begin{table}
\begin{center}
\caption{The number of terms in $a, \lambda, (K,H),(P,Q)$ used for four 
separate solutions of Saturn's orbit based on the Bretagnon model.}
\label{table: bret_model}
\vspace{0.5cm}
\begin{tabular}{|c|c|c|c|c|}
\hline
 &  &  &  & \\ 
Model & $a$ & $\lambda$ & $(K,H)$ & $(P,Q)$ \\
 &  &  &  & \\ 
\hline
 &  &  &  &  \\
SEC  &  1  & 12  & 2 & 2 \\
A21SEC & 12 & 12 & 2 & 2 \\
GISEC & 1 & 12 & 8 & 6 \\
A21GISEC & 12 & 12 & 8 & 6 \\
 &  &  &  &  \\ 
\hline
\end{tabular}
\end{center}
\end{table}
\end{small}

\clearpage

\begin{small}
\begin{table}
\begin{center}
\caption{Results of a ${\chi}^2$ test that compares the 
filtered results of four Bretagnon simulations with the filtered results 
for a full simulation involving Saturn and Jupiter.
Smaller values of ${\chi}^2$ indicates a better fit (the maximum number 
of degree of freedom is 399, see appendix).  The models are 
described in the text.}
\label{table: bret_res}
\vspace{0.5cm}
\begin{tabular}{|c|c|c|c|c|}
\hline
 &  &  &  &  \\ 
Model & SEC & A21SEC & GISEC & A21GISEC \\
 &  &  &  &  \\ 
\hline
 &  &  &  &  \\
${\chi}^2$ & 2741.6 & 132.3 & 70.7 & 33.5 \\
 &  &  &  &  \\ 
\hline
\end{tabular}
\end{center}
\end{table}
\end{small}

\clearpage

\begin{small}
\begin{table}
\begin{center}
\caption{Values of the asymmetry coefficient for four 
simulations with different initial conditions, corresponding 
to particles all having the same values of a) 
$\Omega-{\Omega}_{\odot}$ = constant, b) $\varpi-{\varpi}_{\odot} = 
0^{\circ}$, 
c) $\varpi-{\varpi}_{\odot} = 180^{\circ}$, 
$\lambda-{\lambda}_{\odot} =180^{\circ}$, and 
d) $2(\varpi-{\lambda}_{\odot}) = 45^{\circ}$.  
Case b) seems to be the most symmetric, but all results
overlap to within the errors.}
\label{table: asy_res}
\vspace{0.5cm}
\begin{tabular}{|c|c|c|c|}
\hline
  &  &  &  \\ 
Nodal Resonance & Pericentric Resonance & Pericentric Resonance 
& Evection Inequality\\
$\Omega-{\Omega}_{\odot}$ = const. & $\varpi-{\varpi}_{\odot} = 0^{\circ}$ &
$\varpi-{\varpi}_{\odot} = 180^{\circ}$ & $2(\varpi-{\lambda}_{\odot}) = 
45^{\circ}$ \\
  &  &  &  \\ 
\hline
  &  &  &  \\
60$\pm$13\% & 55$\pm$12\% & 64$\pm$14\% & 60$\pm$18\% \\
  &  &  &  \\ 
\hline
\end{tabular}
\end{center}
\end{table}
\end{small}

\clearpage

\begin{small}
\begin{table}
\begin{center}
\caption{Resonant arguments for the secondary resonances we observed or 
suppose to exist for the region near the separatrix of the Kozai resonance.
The second column reports if the resonant argument was observed to librate.
We used the word ``suspected'' or ``strongly suspected'' for 
resonances inside the libration region, whose positions, when the period
of the Great Inequality was changed, moved according to our predictions.
Other resonances involving higher commensurabilities between $\omega$ 
($=\varpi-\Omega$) and the Great Inequality are weaker, and are not
reported.}
\label{table: Resonant Arguments}
\vspace{0.5cm}
\begin{tabular}{|c|c|}
\hline
  &  \\ 
Resonant argument & Observed \\
  &  \\
\hline
  &  \\
 $\varpi-\Omega-5\lambda_{S}+2\lambda_{J}+3\Omega_{S}$ &  yes \\
 $\varpi-\Omega-5\lambda_{S}+2\lambda_{J}+2\varpi_{S}+\Omega_{S}$ &  yes \\
 $\varpi-\Omega-5\lambda_{S}+2\lambda_{J}+2\varpi_{J}+\Omega_{S}$ &  no \\
 $3(\varpi-\Omega)-2(5\lambda_S-2\lambda_J)+5\Omega_S+\varpi_S$ &  no \\
 $3(\varpi-\Omega)-2(5\lambda_S-2\lambda_J)+5\varpi_S+\Omega_S$ &  no \\
 $3(\varpi-\Omega)-2(5\lambda_S-2\lambda_J)+5\varpi_J+\Omega_S$ &  no \\
 $4(\varpi-\Omega)-3(5\lambda_S-2\lambda_J)+8\Omega_S+\varpi_S$ &  yes \\
 $4(\varpi-\Omega)-3(5\lambda_S-2\lambda_J)+9\varpi_S$ &  yes \\
 $4(\varpi-\Omega)-3(5\lambda_S-2\lambda_J)+9\varpi_J$ &  no \\
 $3(\varpi-\Omega-5\lambda_S+2\lambda_J+3\Omega_S)+(\Omega-\Omega_S)$ & yes \\
 $3(\varpi-\Omega-5\lambda_S+2\lambda_J+3\varpi_S)+(\Omega-\varpi_J)$ & no \\
 $3(\varpi-\Omega-5\lambda_S+2\lambda_J+3\varpi_J)+(\Omega-\varpi_S)$ & no \\
 $2(\varpi-\Omega)-3(5\lambda_S-2\lambda_J)+8\Omega_S+\varpi_S$ &  strongly suspected\\
 $2(\varpi-\Omega)-3(5\lambda_S-2\lambda_J)+9\varpi_S$ &  suspected\\
 $2(\varpi-\Omega)-3(5\lambda_S-2\lambda_J)+9\varpi_J$ &  suspected\\
 $3(\varpi-\Omega)-4(5\lambda_S-2\lambda_J)+11\Omega_S+\varpi_S$ & strongly suspected \\
 $3(\varpi-\Omega)-4(5\lambda_S-2\lambda_J)+11\varpi_S+\Omega_S$ & suspected \\
 $3(\varpi-\Omega)-4(5\lambda_S+2\lambda_J)+11\varpi_J+\Omega_S$ & suspected \\
 & \\
\hline
\end{tabular}
\end{center}
\end{table}
\end{small}

\end{document}